\documentclass[draft]{agujournal2019}
\usepackage{url} 
\usepackage{lineno}
\usepackage[final]{trackchanges} 
\usepackage{soul}

\draftfalse

\def\dbdt{d$B$/d$t$}
\def\dbxdt{d$B_x$/d$t$}
\def\dbydt{d$B_y$/d$t$}
\def\thxn{$\theta_{x_n}$}
\def\thbn{$\theta_{B_n}$}
\def\phiyn{$\varphi_{y_n}$}
\def\RE{$R_E$}
\def\XMP{$X_{mp}$}

\journalname{Space Weather}

\begin{document}

\title{Impact angle control of local intense \dbdt{} variations during shock-induced substorms} 

\authors{Denny M. Oliveira\affil{1,2}, 
		 James M. Weygand\affil{3}, 
		 Eftyhia Zesta\affil{2},
		 Chigomezyo M. Ngwira\affil{4}, 
		 Michael D. Hartinger\affil{5},
		 Zhonghua Xu\affil{6,7,8},
		 Barbara L. Giles\affil{2},
		 Dan J. Gershman\affil{2},
		 Marcos V. D. Silveira\affil{9},
		 V\'itor M. Souza\affil{9}
		 }

\affiliation{1}{Goddard Planetary Heliophysics Institute, University of Maryland, Baltimore County, 
				Baltimore, MD, USA}
\affiliation{2}{Heliophysics Science Division, NASA Goddard Space Flight Center, Greenbelt, MD, USA}
\affiliation{3}{Department of Earth, Planetary, and Space Sciences, University of California Los Angeles, 
				Los Angeles, CA, USA}
\affiliation{4}{Atmospheric and Space Technology Research Associates LLC, Louisville, CO, USA}
\affiliation{5}{Space Science Institute, Boulder, CO, USA}
\affiliation{6}{Center for Space Science and Engineering Research, Virginia Tech, Blacksburg, VA, USA}
\affiliation{7}{Bradley Department of Electrical and Computer Engineering, Virginia Tech, Blacksburg, VA, USA}
\affiliation{8}{National Institute of Aerospace, Hampton, VA, USA}
\affiliation{9}{National Institute for Space Research – INPE, S\~ao Jos\'e dos Campos, SP, Brazil}

\correspondingauthor{Denny Oliveira}{denny@umbc.edu}

\begin{keypoints}
	
	\item We perform a comparative study of geospace and ground multi-instrument response to two similarly strong shocks with different orientations
	
	\item Energetic particle injections and \dbdt{} variations are faster and more intense in the nearly frontal case due to symmetric compression
	
	\item Areas with intense \dbdt{} peaks are larger and occur earlier in the first case due to the symmetric compression by the head-on shock

\end{keypoints}

\begin{abstract}

	The impact of interplanetary shocks on the magnetosphere can trigger magnetic substorms that intensify auroral electrojet currents. These currents enhance ground magnetic field perturbations (\dbdt), which in turn generate geomagnetically induced currents (GICs) that can be detrimental to power transmission infrastructure. We perform a comparative study of \dbdt{} variations in response to two similarly strong shocks, but with one being nearly frontal, and the other, highly inclined. Multi-instrument analyses by the Time History of Events and Macroscale Interactions during Substorms (THEMIS) and Los Alamos National Laboratory spacecraft show that nightside substorm-time energetic particle injections are more intense and occur faster in the case of the nearly head-on impact. The same trend is observed in \dbdt{} variations recorded by THEMIS ground magnetometers. THEMIS all-sky imager data show a fast and clear poleward auroral expansion in the first case, which does not clearly occur in the second case. Strong field-aligned currents computed with the spherical elementary current system (SECS) technique occur in both cases, but the current variations resulting from the inclined shock impact are weaker and slower compared to the nearly frontal case. SECS analyses also reveal that geographic areas with \dbdt{} surpassing the thresholds 1.5 and 5 nT/s, usually linked to high-risk GICs, are larger and occur earlier due to the symmetric compression caused by the nearly head-on impact. These results, with profound space weather implications, suggest that shock impact angles affect the geospace driving conditions and the location and intensity of the subsequent \dbdt{} variations during substorm activity.

\end{abstract}

\section*{Plain Language Summary}
	
	Solar perturbations propagating in the interplanetary space can cause significant geomagnetic activity when they impact Earth. Such magnetic disturbances occur in the geospace and on the ground, manifested as, e.g., satellite surface charging and undesirable geoelectric currents flowing in large-scale power transmission lines. In this study, we compare the effects caused by two similarly strong solar perturbations that impacted Earth with two very distinct orientations: one nearly head-on, and the other highly inclined. We use an extensive list of data sets covering observations in the interplanetary space, in the geospace, and on the ground. We find that magnetic field perturbations and auroral brightening are much more intense in the nearly frontal case, while these observations are considerably weaker in the second case. We attribute these effects to the highly symmetric conditions in which electromagnetic energy was released in the first case, which were quite asymmetric in the second case. These results are relevant to scientists whose goal is to predict and forecast geomagnetic activity that will follow solar perturbations detected in the interplanetary space by solar wind monitors.

\section{Introduction}

	Interplanetary (IP) shocks are solar wind structures commonly observed in the heliosphere \cite{Winslow2015,Mihalov1987,Harada2017,Echer2019,Richardson2005}. IP shocks arise when the ratio of the relative speed (between the ambient solar wind and the shock itself) to the local magnetosonic speed, defined as the magnetosonic Mach number, is greater than 1 \cite{Jeffrey1964,Priest1981}. The most common and geoeffective type of shocks is named fast forward shock, which propagates away from the Sun \cite{Tsurutani2011a,Oliveira2017a}. The strength of an IP shock is commonly measured by magnetosonic Mach numbers, and ratios of shocked to unshocked solar wind number densities and ram pressures \cite{Tsurutani1985b,Andreeova2007b,Kwon2018}. \par 

	The first prompt and immediate magnetospheric response to shock impacts is characterized as sharp variations in the geomagnetic field observed in space \cite{Kokubun1983,Wang2009,Villante2011,Bhaskar2021}, and on the ground \cite{Chao1974,Araki1994,Veenadhari2012,Smith2020a}. IP shocks are also known to trigger substorm activity \cite{Burch1972,Akasofu1980,Zhou2001,Meurant2004}. Substorms are triggered by the explosive release of energy resulting from magnetospheric instabilities \cite{Wing2014,Kalmoni2018}. The occurrence of substorms is commonly associated with high intensifications of auroral westward currents \cite{McPherron1993,Lyons1996,Newell2011a}, which are commonly quantified by westward auroral electrojet indices, such as the SML index \cite<>[see also section 2.3]{Orr2021}. Supersubstorms are usually classified as events with minimum SML $\leq$ --2500 nT, and intense substorms are events with minimum SML in the interval --2500 nT $<$ SML $\leq$ --2000 nT \cite<e.g.,>[]{Tsurutani2021}. Such events are usually isolated and weakly correlated with magnetic storms \cite{Tsurutani2015,Tsurutani2021,Zong2021}. Preconditions defined by negative values of the southward component of the interplanetary magnetic field (IMF), fast solar wind, and positive Russell-McPherron effect determine whether substorms are triggered or not by shocks \cite{Russell1973b,Craven1986,Zhou2001,Weygand2014,Zong2021}. \par

	One well-known and important space weather effect detected on the ground is manifested as the so-called geomagnetically induced currents (GICs). Sharp variations of the surface geomagnetic field (\dbdt) generate geoelectric fields which in turn couple with artificial conductors giving rise to the flow of undesirable GICs \cite{Viljanen1998,Oliveira2017d,Belakhovsky2019,Abda2020}. Although ground conductivity models are necessary to compute geoelectric fields that generate GICs \cite{Love2016b,Liu2019a}, \dbdt{} variations are an acceptable and widely used proxy for assessing GIC response to space weather because such geoelectric fields may exist independent of ground artificial conductors. Therefore, in this work, the focus will be on \dbdt{} variations. dB/dt variations are usually more intense and frequent at high latitudes during storm-time substorms \cite{Bolduc2002,Wik2008,Pulkkinen2012}, but significantly intense dB/dt variations can also be observed at mid- and low-latitudes following impacts of IP shocks \cite{Fiori2014,Carter2015,Oliveira2018b,Keesee2020}. Power grid equipment and transmission lines can be subjected to high risks when dB/dt surpasses the thresholds 1.5 nT/s \cite{Pulkkinen2013} and 5 nT/s \cite{Molinski2000}. \citeA{Weygand2021} showed that dB/dt variations for these thresholds are quite common during non-storm-time substorms. According to \citeA{Zong2021}, \dbdt{} variations triggered by supersubstorms are likely to shift energetic particle injections inward and accelerate particles in an existing population. \par

	Recently, \citeA{Ngwira2018c} performed a careful and detailed study of the magnetosphere-ionosphere driving conditions that led high-latitude magnetometers to observe intense local \dbdt{} variations during two intense magnetic storms. The authors reported that such variations can occur shortly after intense and rapid substorm-time energetic particle injections that were mapped back to observations performed by multi-instruments on board spacecraft in the magnetosphere. \citeA{Ngwira2018c} also noted, based on auroral image observations, that the substorm onset and rapid poleward auroral oval expansion were shortly followed by intense \dbdt{} variations. In addition, the authors showed that during these storm-time substorms intense ultra-low frequency (ULF) wave activity may have further added more complexity to the system response indicating that wave activity also contributed to the ground \dbdt{} variations. \par  

	Shock normal orientations determine how shocks propagate in the heliosphere. Consequently, when shocks hit Earth, different levels of geomagnetic activity may follow, including ground sudden impulses, field-aligned current and auroral intensifications, and wave response in the magnetosphere-ionosphere system \cite<e.g., see review by>{Oliveira2018a}. Observations and simulations show that geomagnetic activity following the impact of an inclined shock is usually weaker and slower than the geomagnetic activity following a frontal shock as a result of asymmetric magnetospheric compressions \cite{Takeuchi2002b,Guo2005,Grib2006,Wang2006a,Samsonov2011a,Samsonov2015,Selvakumaran2017,Shi2019b,Xu2020a}. Numerical simulations conducted by \citeA{Oliveira2014b} showed that a frontal shock triggered intense substorm activity, whereas an inclined shock triggered moderate substorm activity, even though the inclined shock was stronger. In a statistical study with data provided by a worldwide array of ground magnetometers, \citeA{Oliveira2015a} showed that shock impact angles correlated remarkably well with westward auroral electrojet currents.

	Therefore, the goal of this paper is to build on the work of \citeA{Ngwira2018c} and establish the first clear observational link between shock impact angles and the subsequent \dbdt{} intensifications during isolated substorms. We compare \dbdt{} variations following the impact of two similarly strong shocks, but with very different normal orientations. We shall demonstrate that the almost head-on shock impact led ground magnetometers to register more frequent and more intense \dbdt{} variations covering larger geographic areas in comparison to the inclined shock.

\section{Data sets}

	\subsection{Solar wind and interplanetary magnetic field data}

		We use solar wind and IMF data from the Wind and Advanced Composition Explorer (ACE) spacecraft located in the solar wind ahead of the Earth at the Lagrangian point L1. Wind solar wind velocity, number density, and temperature data are provided by the Solar Wind Experiment instrument \cite{Ogilvie1995}, whereas the IMF data are provided by the Magnetic Fields Investigation instrument \cite{Lepping1995}. ACE plasma parameter and IMF data are provided by the Solar Wind Electron, Proton, and Alpha Monitor instrument \cite{McComas1998}, and by the Magnetic Fields Experiment instrument \cite{Smith1998}, respectively. The solar wind and IMF data used in this study have resolution of 1 minute. \par

	\subsection{Geospace-based plasma and magnetic field data}

		Geospace plasma and magnetic field data collected by the Time History of Events and Macroscale Interactions during Substorms (THEMIS) mission are used in this study. The primary purpose of the initial five THEMIS spacecraft is to investigate the time evolution and generation mechanisms of magnetospheric substorms using conjugated ground-based and space-borne observations \cite{Angelopoulos2008a,Sibeck2008}. The Electrostatic Analyzer (ESA) and the Solid State Telescope (SST) on board THEMIS measure plasma data within a 3-second cadence. Magnetic field data are sampled by the Flux Gate Magnetometer (FGM) instrument 64 times per second. The ESA and FGM instruments are described with details by \citeA{McFadden2008} and \citeA{Auster2008}, respectively. \par

		\begin{figure}[t]
			\centering
			\includegraphics[width=14cm]{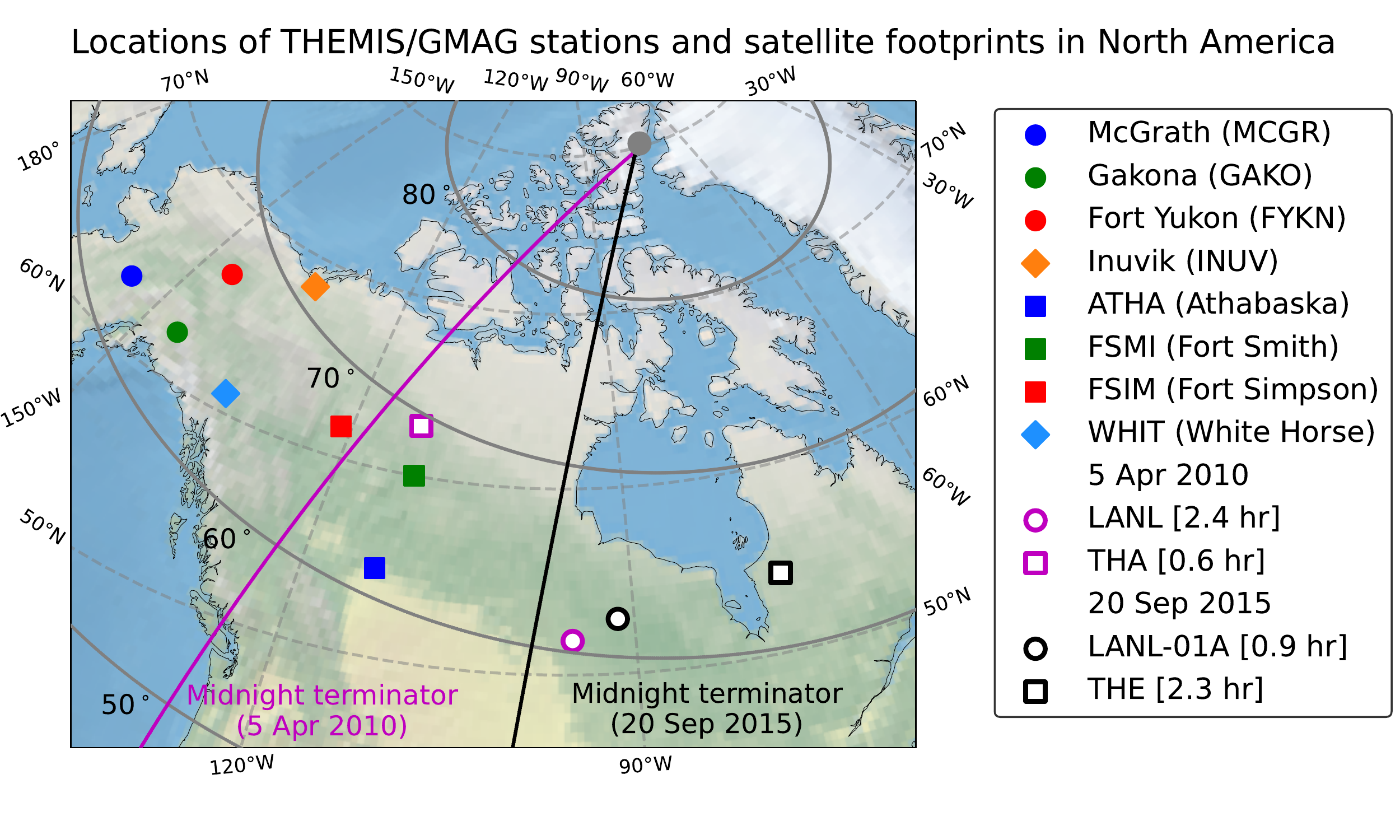}
			\caption{Geographic locations of the THEMIS/GMAG stations in northwestern North America used in this study. Colored circles represent stations whose data are used for the nearly frontal shock, while colored squares represent stations whose data are used for the highly inclined shock. Colored diamonds represent stations whose data are used for both shocks. The white circles and squares with magenta and black edges show the magnetic footprints of satellites in geospace at each corresponding substorm onset. The thick solid magenta and black lines indicate the position of the magnetic local midnights for the nearly frontal shock and the highly inclined shock, respectively.}
			\label{stations}
		\end{figure}

		In geosynchronous orbit, we use data provided by Los Alamos National Laboratory (LANL) spacecraft \cite{Belian1992,Thomsen1999}. We particularly use electron flux data collected by the Synchronous Orbit Particle Analyzer (SOPA) from the entire three-dimensional electron distribution with energies ranging from 50 keV to over 1.5 MeV in each spin. The resolution of the SOPA electron flux data is 10 s \cite{Thomsen1999}. \par

		Magnetospheric Multiscale \cite<MMS>[]{Burch2016a} observations of geospace plasma and magnetic field are also used. The MMS Flux Gate Magnetometer (FGM) slow survey data cadence is 8 measurements per second \cite{Torbert2014,Russell2016b}, while the MMS Fast Plasma Investigation \cite<FPI>[]{Pollock2016} cadences are 150 and 30 ms for ions and electrons, respectively.

	\subsection{Ground-based magnetic field and image data}

		Geomagnetic activity triggered by IP shocks in this work is expressed by SuperMAG index data \cite{Gjerloev2009}. Ring current effects are represented by the SuperMAG ring current (SMR) index \cite{Newell2012}. The SuperMAG auroral indices, represented by SMU, SML, and SME, are similar to the standard AU, AL, and AE indices \cite{Davis1966}, but more ground magnetometer stations are used by the SuperMAG array for index computations \cite{Newell2011a}. All SuperMAG data used in this study are 1-minute resolution data. We use the SuperMAG dataset because the SuperMAG indices are usually more likely to capture geomagnetic activity triggered by shocks due to their larger and more effective geographic coverage \cite<e.g.>{Oliveira2016a,Oliveira2015a,Oliveira2018a,Rudd2019,Shi2019b,Oliveira2020d}. \par

		\begin{table}
			\centering
			\begin{tabular}{l c c c c c}
				\hline
					\multicolumn{6}{c}{5 April 2010/0903 UT (Nearly frontal shock)} \\
					Station name/code & GLAT [$^\circ$] & GLON [$^\circ$] & MLAT [$^\circ$] & MLON [$^\circ$] & MLT [hrs] \\
				\hline
					McGrath (MCGR)  	& 63.0  &  --155.6  &   62.4  &  --102.8  &   21.4 \\
					Gakona (GAKO)		& 62.4  &  --145.2  &   63.7  &   --92.6  &   22.1 \\
					White Horse (WHIT)	& 61.0  &  --135.2  &   64.1  &   --81.4  &   22.8 \\
					Fort Yukon (FYKN)	& 66.6  &  --145.3  &   67.5  &   --96.7  &   21.8 \\
					Inuvik (INUV)		& 68.3  &  --133.3  &   71.1  &   --87.3  &   22.4 \\
				\hline
				\multicolumn{6}{c}{20 September 2015/0703 UT (Highly inclined shock)} \\
					Station name/code & GLAT [$^\circ$] & GLON [$^\circ$] & MLAT [$^\circ$] & MLON [$^\circ$] & MLT [hrs] \\
				\hline
					Athabaska (ATHA)	& 54.7  &  --113.3  &   61.4  &   --51.9  &   22.7	\\
					White Horse (WHIT)	& 61.0  &  --135.2  &   64.1  &   --80.3  &   20.9	\\
					Fort Smith (FSMI)	& 60.0  &  --111.9  &   66.7  &   --53.2  &   22.7	\\
					Fort Simpson (FSIM)	& 61.8  &  --121.2  &   67.1  &   --65.6  &   21.8 \\
					Inuvik (INUV)		& 68.3  &  --133.3  &   71.1  &   --86.0  &   20.5	\\
				\hline
			\end{tabular}
			\caption{Geographic and magnetic coordinates of the 8 THEMIS/GMAG stations used in this study. Magnetic coordinates are computed by the centered-dipole model around the substorm onsets at 0903 UT (nearly frontal shock) and 0703 UT (highly inclined shock) on 5 April 2010 and 20 September 2015, respectively.}
			\label{table1}
		\end{table}

		We use THEMIS ground magnetometer (GMAG) data \cite{Russell2008b,Mende2008} to investigate ground magnetic field response during nighttime substorms triggered by shocks. We also use white-light THEMIS All-Sky Imager (ASI) data \cite{Donovan2006} to support the identification of auroral substorm activity and onset occurrence. The THEMIS/GMAG and ASI array contains stations spread across North America from western Alaska to eastern Labrador \cite{Donovan2006,Russell2008b}. As described by \citeA{Russell2008b}, the main objective of THEMIS/GMAG is to provide support to THEMIS observations in space. Although there are over 30 GMAGs and 20 ASIs available in the THEMIS array, we use only eight distinct GMAGs and ASIs in our study because these stations were located near the magnetic local midnight during substorm onsets. This region corresponds to western North America. The THEMIS/GMAG magnetic field data resolution is either 1 second or 0.5 second \cite{Russell2008b}, but all stations used in this study performed 2 observations per second. The ASI cameras take white light images with cadence of 3 seconds in a 256$\times$256 pixel image resolution \cite{Donovan2006}. Figure \ref{stations} shows a North American map with the THEMIS/GMAG and ASI stations, and Table \ref{table1} shows geographic and geomagnetic information (magnetic latitude, MLAT; magnetic longitude, MLON; and magnetic local time, MLT) for these stations. In this work, magnetic coordinates are computed with the centered-dipole (CD) model \cite{Emmert2010,Laundal2017} and the 13th generation of the International Geomagnetic Reference Field \cite{Alken2021}

\section{Definition and computation of shock parameters}

	\subsection{Shock normal vector and speeds}

		IP shocks are assumed to be planar structures when traveling in the heliosphere at 1 AU. Therefore, the shock normal vector can be computed for all shocks if data from at least one spacecraft are available \cite{Russell1983}. The Rankine-Hugoniot conditions, obtained from energy and momentum conservations across the shock surface \cite{Vinas1986,Schwartz1998,Oliveira2017a}, are employed for the computation of shock normal coordinates and speeds. Specifically, we use the following equation that combines solar wind and IMF data for shock normal determinations:

		\begin{equation}
			\vec{n} = \frac{(\Delta\vec{B}\times\Delta\vec{v})\times\Delta\vec{B}}{|(\Delta\vec{B}\times\Delta\vec{v})\times\Delta\vec{B}|}\,,
		\end{equation}
		where $\Delta \vec{B} = \vec{B}_2 - \vec{B}_1$, $\Delta \vec{v} = \vec{v}_2 - \vec{v}_1$, and the indices (1) and (2) represent the upstream and downstream regions, respectively. Therefore, this equation yields \thxn{} = $\cos^{-1}(n_x)$, the angle the normal vector performs with the geocentric solar ecliptic (GSE) x-axis, and \phiyn = $\tan^{-1}(n_z/n_y)$, the angle between the shock normal vector and the GSE y-axis in the GSE yz plane. \par

		Shock strengths are usually defined by the following shock parameters. The shock speed is computed as $v_s=(\rho_2\vec{v}_2 - \rho_1\vec{v}_1)\cdot\vec{n}/(\rho_2-\rho_1)$, where $\rho_1,\rho_2$ are the upstream, downstream solar wind densities, respectively. The compression ratio is computed as $X_\rho=\rho_2/\rho_1$. The solar wind dynamic pressure ratio is given by $X_{P_d}=(\rho_2v_2^2)/(\rho_1v_1^2)$. The fast magnetosonic speed is given by $v_{ms} =\sqrt{1/2[(v_A^2 + c_s^2) + \sqrt{(v_A^2 + c_s^2)^2 - 4v_A^2c_s^2\cos^2\theta_{B_n}}]}$ with the Alfv\'en speed given by $v_A=|\vec{B_1}|/\sqrt{\mu_0\rho_1}$, where $\mu_0$ represents the magnetic permeability in free space, the sound speed given by $c_s = \sqrt{\gamma P_1 / \rho_1}$, with $\gamma$ representing the ratio of the plasma specific heat with constant pressure to the plasma specific heat with constant volume ($\gamma$ = 5/3), $P_1$, upstream plasma thermal pressure, and \thbn{}, the angle between the upstream magnetic field vector and the shock normal vector. Therefore, the magnetosonic Mach number is given by $M_{ms}=v_r/v_{ms}$, where $v_r=v_s-v_1$.

	\subsection{Shock impact angle and speed conventions}

		According to equation (1), the angle \thxn{} = 180$^\circ$ indicates a purely frontal shock. When \thxn{} decreases within the interval 90$^\circ$ $<$ \thxn{} $<$ 180$^\circ$, the shock becomes inclined \cite{Oliveira2017a,Oliveira2018a}. On the other hand, a shock whose impact angle is in the interval 0$^\circ$ $<$ \thxn{} $<$ 90$^\circ$ in GSE coordinates is classified as a reverse shock. In the same reference system, \thxn{} = 0$^\circ$ indicates a purely frontal reverse shock. Therefore, the expression ``small impact angle" refers to shocks whose normals are almost aligned with the Sun-Earth line. The shock speed $v_s$ is measured by a satellite in the solar wind at L1 with respect to Earth or the satellite itself \cite{Colburn1966,Tsurutani2011a,Oliveira2017a}.

		\begin{table}
			\centering
			\begin{tabular}{l c c}
				\hline
					Date/UT 			& 5 Apr 2010/0826 & 20 Sep 2015/0603 \\
					Satellite 			& ACE 		   		& Wind 			   			 \\
					Inclination 		& NFS 				& HIS 						\\
				\hline
					\thxn{} [$^\circ$]	& 171.7		   		& 143.2			   			 \\
					\thbn{} [$^\circ$]	& 38.1		   		& 47.8			   			 \\
					\phiyn{} [$^\circ$]	& 130.4		   		& 17.6						 \\
					$v_s$ [km/s]		& 796.6				& 615.5						 \\
					$M_{ms}$ 				& 2.0				& 2.6 						\\
					$X_\rho$ 			& 2.70 				& 1.71						\\
					$X_{P_d}$			& 4.31				& 2.55 						\\
					\XMP{} [\RE]		& 5.39				& 6.43 						\\
					$B_{z2}$ [nT]			& --12 				& --14						\\			
				\hline
			\end{tabular}
			\caption{Shock parameters computed by the Rankine-Hugoniot conditions (equation (1)) with solar wind and IMF data: \thxn, shock impact angle; \thbn, angle between upstream magnetic field and shock normal vector; \phiyn, the angle between the shock normal vector and the GSE y-axis in the GSE yz plane; $v_s$, shock speed; $M_{ms}$, magnetosonic Mach number; $X_{\rho}$ downstream to upstream solar wind density ratio; $X_{P_d}$ downstream to upstream solar wind ram pressure ratio; \XMP, magnetopause standoff position; and $B_{z2}$, the average of the z component of the IMF in the two hours following each respective shock impact.}
			\label{table2}
		\end{table}

\section{Interplanetary shock observations in the solar wind}
	
	\subsection{Conditions for event selection}

		This study focuses on geoeffectiveness comparisons between two IP shocks: a nearly frontal shock (NFS, \thxn{} $\sim$ 172$^\circ$), observed by ACE on 5 April 2010, and a highly inclined shock (HIS, \thxn{} $\sim$ 143$^\circ$), observed by Wind on 20 September 2015. Both shocks were driven by coronal mass ejections (CMEs). According to the CME list (\url{http://www.srl.caltech.edu/ACE/ASC/DATA/level3/icmetable2.htm#(a)}) provided and maintained by \citeA{Richardson2010b}, the mean transient speeds of the 2010 and 2015 CMEs were 910 and 850 km/s, respectively, meaning that these CMEs can be classified as moderately fast in the context of solar cycle 24 \cite{Ravishankar2019}. Table \ref{table2} summarizes the upstream and downstream conditions of both shocks. These shocks were selected because they had similar strengths, as shown by the parameters in Table \ref{table2}, but they had very distinct inclinations. The inclination effects are evident in the table as shown by the ratios of downstream to upstream densities and ram pressures, which are higher for the NFS due to symmetric compressions \cite<e.g.,>[]{Oliveira2018a}. Additionally, as we will show later, these shocks were selected because they triggered isolated substorms with significant intensities. The events were selected from the shock list provided by \citeA{Oliveira2018b}. \par

		\begin{figure}[t]
			\centering
			\includegraphics[width=14cm]{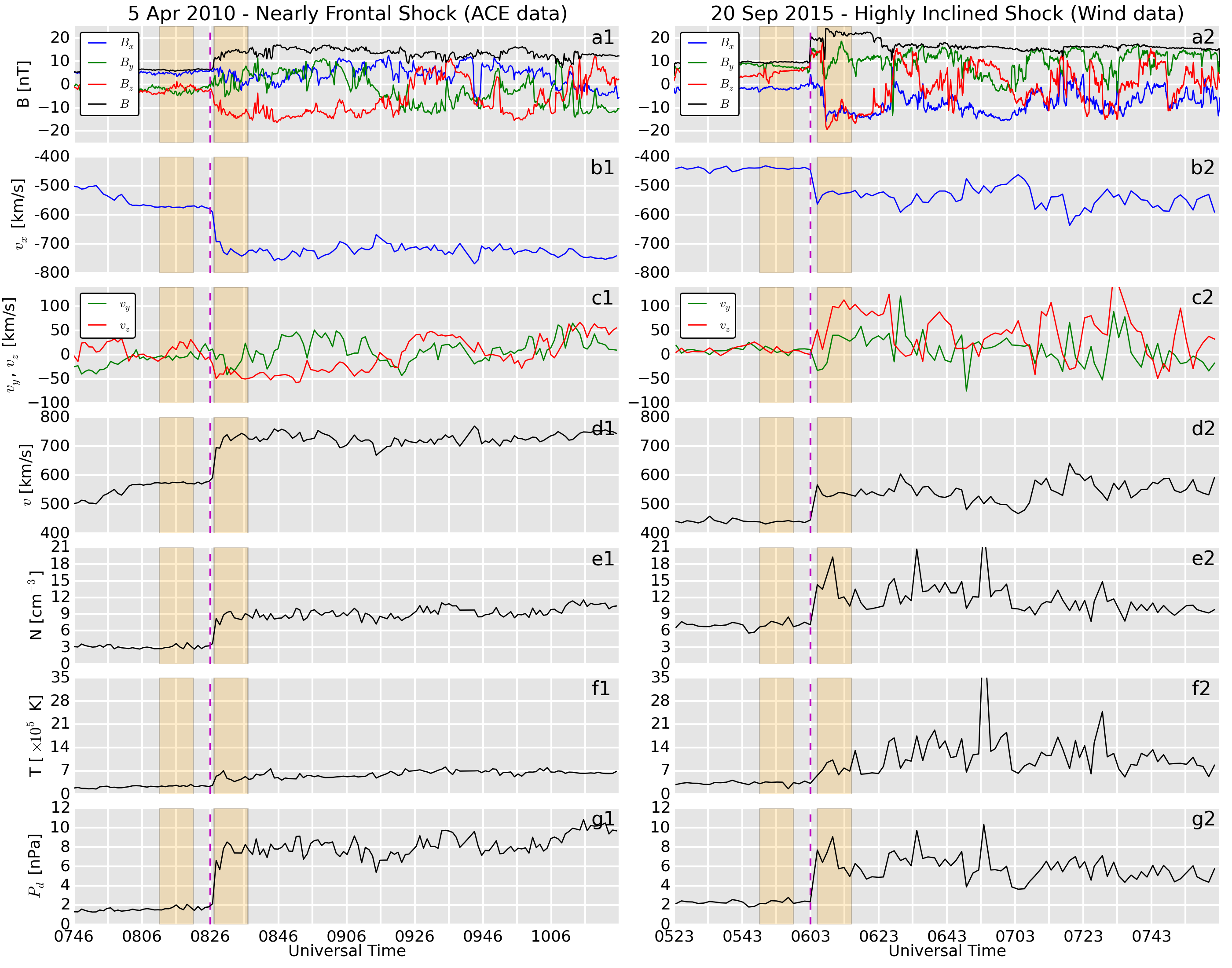}
			\caption{Summary plots of IMF and solar wind conditions for the shocks observed at L1 ahead of the Earth by ACE (5 April 2010, left column) and Wind (20 September 2015, right column). The dashed vertical purple lines indicate the respective shock onset. The light orange highlighted areas correspond to the upstream (before shock) and downstream (after shock) for each respective satellite. In comparison to the frontal case, the downstream window for the inclined case is moved 2 minutes ahead because inclined shocks usually take more time to form.}
			\label{shocks}
		\end{figure}

		Figure \ref{shocks} shows solar wind plasma parameter and IMF observations for the NFS (left column) and the same for the HIS (right column). The data, shifted to the respective magnetopause position, are plotted within the interval (--40, 120) minutes around each respective $t_0$ (0826 and 0603 UT), the shock onset time (dashed purple lines). The highlighted areas before and after the shock represent the upstream and downstream regions selected for use in equation (1). The upstream and downstream windows were carefully chosen as [(--15, --5), (1, 11)] minutes and [(--15, --5), (3, 13)] minutes for both shocks, respectively. Note that the downstream HIS window was moved 2 minutes ahead in comparison to NFS because the HIS took more time to develop in comparison to the NFS, as usually occurs with inclined shocks \cite<e.g.,>{Wang2006a,Rudd2019}. The second downstream window was chosen particularly for this event, but it could change arbitrarily in statistical studies with large numbers of inclined shocks. All panels in Figure \ref{shocks} indicate positive jumps in the IMF and plasma parameter magnitudes (velocity, number density, and temperature), which are fast forward shock signatures \cite{Priest1981,Oliveira2017a}. Additionally, the HIS shows more variability in the y and z components of the IMF and solar wind speed due to its normal inclinations with respect to the xy and xz planes \cite{Xu2020a}. Table \ref{table2} also shows that the average southward component of the IMF ($B_{z2}$) was slightly more negative in the downstream region of the HIS (--14 nT) than in the downstream region of the NFS (--12 nT). The average IMF $B_z$ components preceding both shock impacts were nearly null.\par

		The NFS and HIS occurred during the ascending and descending phases of the notably weak solar cycle 24 \cite{Clette2016b}, respectively. Effects of solar activity on the solar wind can be clearly seen in panels g1 and g2 for densities before shock impact (around 3 and 6 cm$^{-3}$), and panels b1 and b2 for solar wind velocity before shock impact ($\sim$580 km/s and $\sim$420 km/s), respectively. Consequently, this translates into similar values for the upstream dynamic pressures, $\sim$2 nPa, which indicates very similar solar wind ram pressure conditions around L1 and similar magnetospheric equilibrium states before shock impacts. However, although both shocks had relatively similar strengths, the HIS took more time to develop reaching smaller peak values in the downstream region, as can be clearly seen in the case of the ram pressure. This is a direct shock impact angle consequence \cite<e.g., >{Oliveira2018a}.

	\subsection{Magnetopause standoff position}

		In order to compare the compression states of the magnetosphere during both shock events, we use the \citeA{Shue1998} empirical model to compute the magnetopause standoff position (\XMP) for each shock as a function of time. The model uses the IMF $B_z$ component and dynamic pressure to account for equilibrium conditions between the magnetopause and the solar wind. \par

		\begin{figure}[t]
			\centering
			\includegraphics[width=13cm]{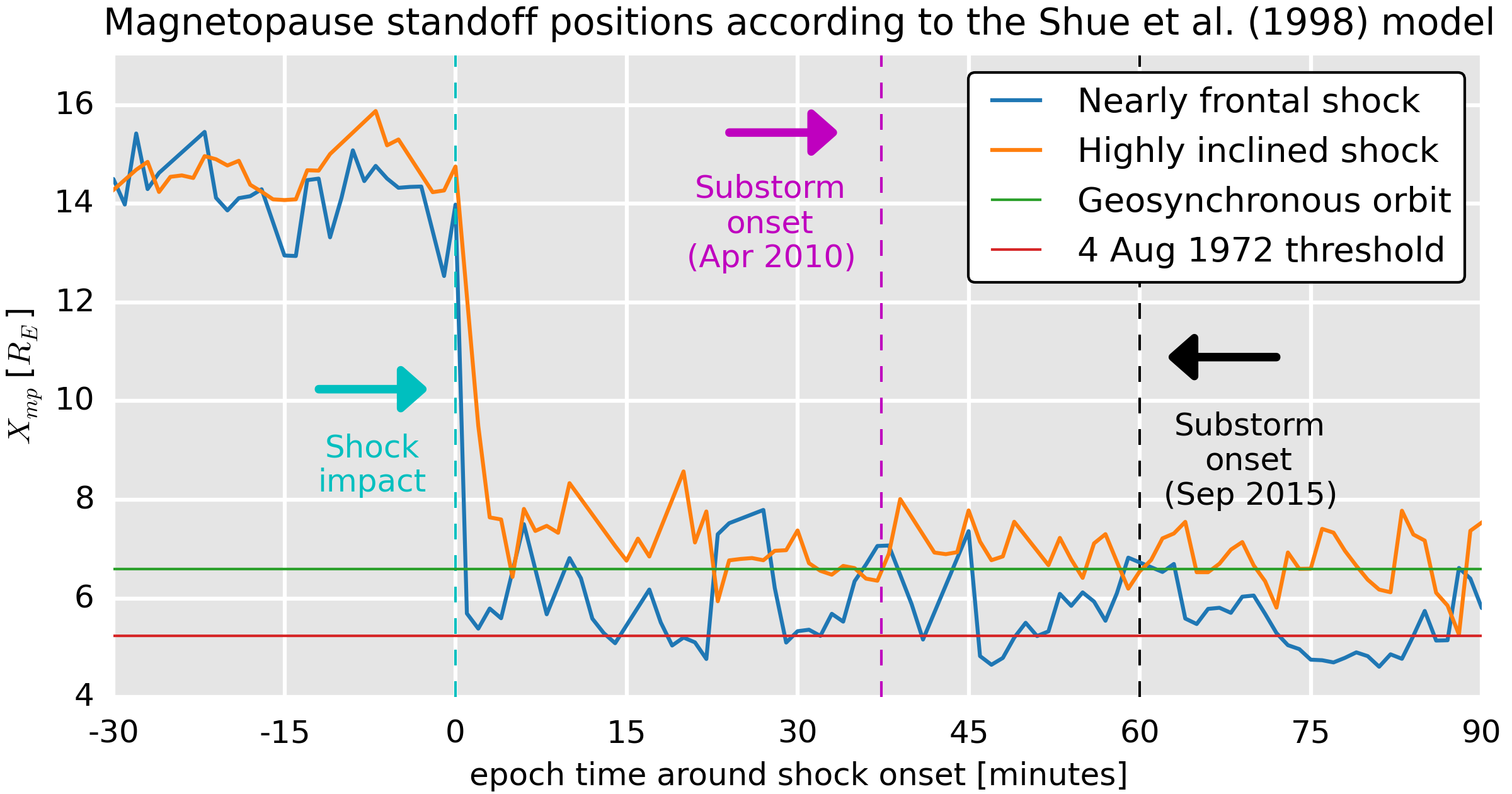}
			\caption{Magnetopause standoff position (\XMP) as a function of epoch time for both shocks: nearly frontal shock (solid blue line), and highly inclined shock (solid orange line). The \citeA{Shue1998} empirical model, based on solar wind and IMF data, was used to compute \XMP.}
			\label{standoff}
		\end{figure}

		Figure \ref{standoff} shows \XMP{} as a function of epoch time (--60, 90) minutes around the corresponding shock onset time. In the figure, the solid blue line is for the NFS, whereas the solid orange line is for the HIS. The model predicts the magnetopause was located at very similar locations prior to the shock impacts, which also indicate similar solar wind dynamic pressure and IMF $B_z$ conditions prior to shock impacts. Results show minimum \XMP{} of 5.4 and 6.5 \RE{} for both shocks, respectively (\RE{} = 6371 km is the Earth's radius). These values are achieved 2 and 5 minutes after shock impacts, the latter consistent with shock inclination effects. Therefore, these observations also indicate that both NFS and HIS were similarly strong events, but with different inclinations. These results are supported by many simulations and observations of shocks impinging the magnetosphere with different orientations \cite{Guo2005,Wang2006a,Oliveira2018a,Shi2019b}. \par 

		The solid horizontal green line in Figure \ref{standoff} indicates the threshold position at geosynchronous orbit ($\sim$6.6\RE). This type of extreme inward motion is a major loss process for radiation belt particles \cite<e.g.>[]{Turner2012}. Satellites may also be exposed to energetic radiation belt particles that may partially or completely damage their electronic systems in the aftermath of this extreme inward motion, when the magnetopause moves back outward and the satellite re-enters the magnetosphere and radiation belts; this damage is more likely to occur when the conditions are right for rapid increases in radiation belt particles, for example the occurrence of particle injections and wave-particle interactions \cite{Koons2006,Baker2017}. The horizontal red line indicates the arguably most inward magnetopause position ever observed, \XMP{} $\approx$ 5.2 \RE, derived by \citeA{Vaisberg1976} from satellite observations during the impact of a very fast CME in August 1972. As can be seen in the figure, \XMP{} surpassed the geosynchronous threshold after both shock impacts with NFS almost reaching the minimum $X_{mp}$ ever observed. Results also show that the magnetosphere was highly compressed due to high dynamic pressure conditions and highly depressed IMF $B_z$ in both events. However, \XMP{} in the NFS case stayed more inward most of the time in comparison to the HIS case due to two main factors (i) stronger downstream dynamic pressure levels, and (ii) highly depressed and steady negative values of IMF $B_z$ due to more symmetric magnetospheric compression by the NSF in comparison to the HIS. These results confirm previous numerical simulations by \citeA{Oliveira2014b} for magnetic field intensifications after shock impacts. \par



\section{Results}

	\subsection{Ground magnetometer response and auroral observations}

\begin{figure*}
			\centering
			\includegraphics[width=16cm]{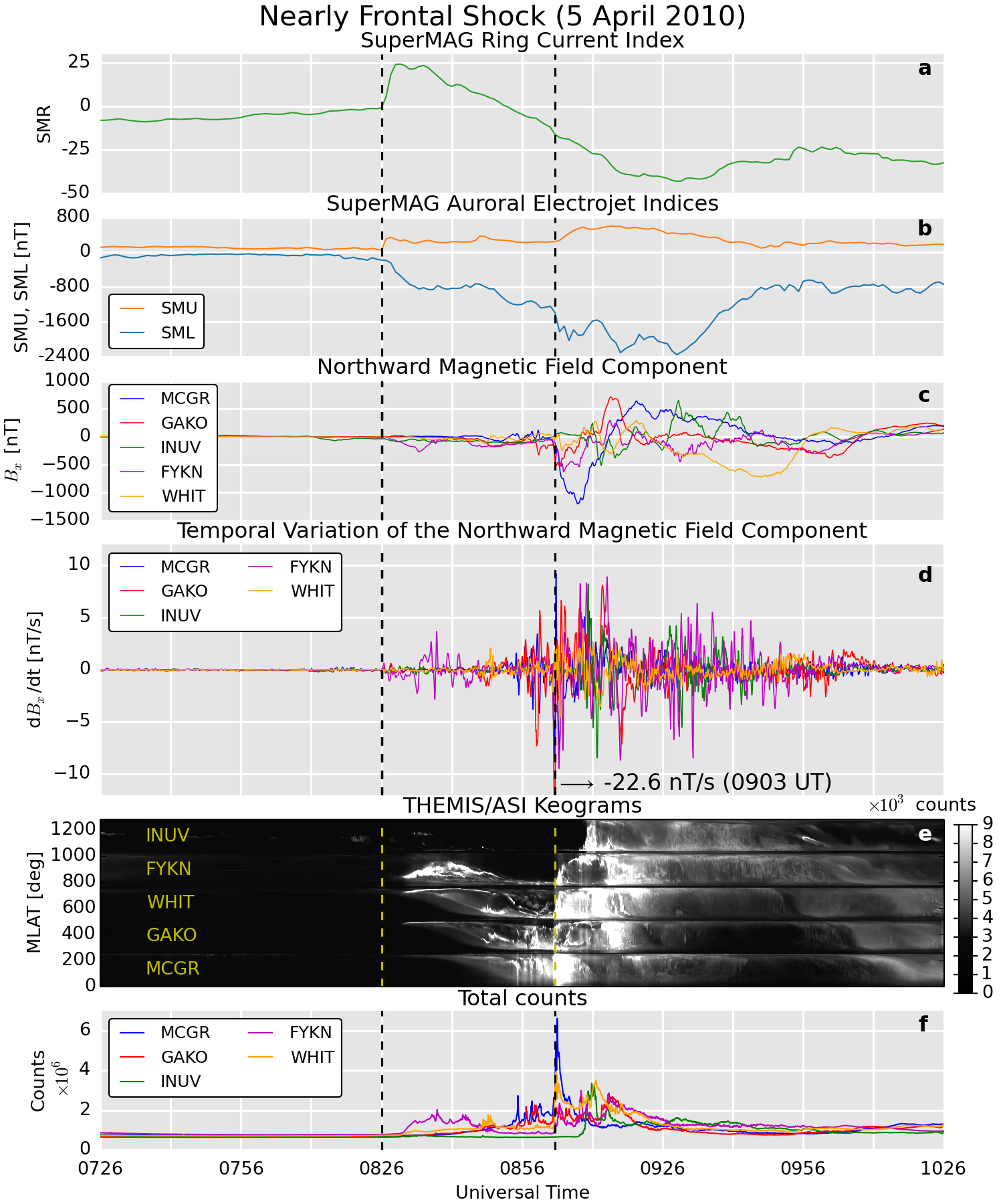}
			\caption{Summary plots of ground magnetometers (SuperMAG and THEMIS, a-d), and THEMIS/ASI (e and f) data for the nearly frontal shock of 5 April 2010.}
			\label{394}
		\end{figure*}

		In this section we focus on the ground magnetic field response and auroral observations resulting from the impact of the shocks. Figure \ref{394} shows results for the NFS (5 April 2010). The figure shows geomagnetic activity data represented by the SuperMAG SMR, SMU, and SML indices (panels a and b); THEMIS/GMAG data for the $B_x$ (northward) magnetic field component (panel c) and its temporal variation \dbxdt{} (panel d); and THEMIS/ASI observations stacked with respect to MLATs (panel e) computed with the CD model. Note that the MLAT separation in this case is not uniform. The sum of counts for each of the 256 pixel bins provided by each individual station is shown in panel f. The quantities plotted in panels d, e, and f are shown with more details for all stations during both events in the supporting information. \par

		A positive sudden impulse (SI$^+$) is shown by the SMR index at 0826 UT, a clear shock impact signature (first vertical dashed line) with a rise time of $\sim$2 min and amplitude $\sim$30 nT. Later, the index dropped to values above --50 nT, indicating moderate ring current activity. Panel b shows intense auroral activity as indicated by the SMU and and SML indices. Low SML values particularly indicate intense activity in the westward auroral electrojet current, a clear substorm signature \cite{Mayaud1980b,Gjerloev2004,Orr2021}. A substorm onset is marked by the second dashed vertical line as indicated by a sharp decrease in SML at 0903 UT from approximately --1200 nT to --1800 nT and later a gradual decrease to --2400 nT. \par

		\begin{figure*}
			\centering
			\includegraphics[width=16cm]{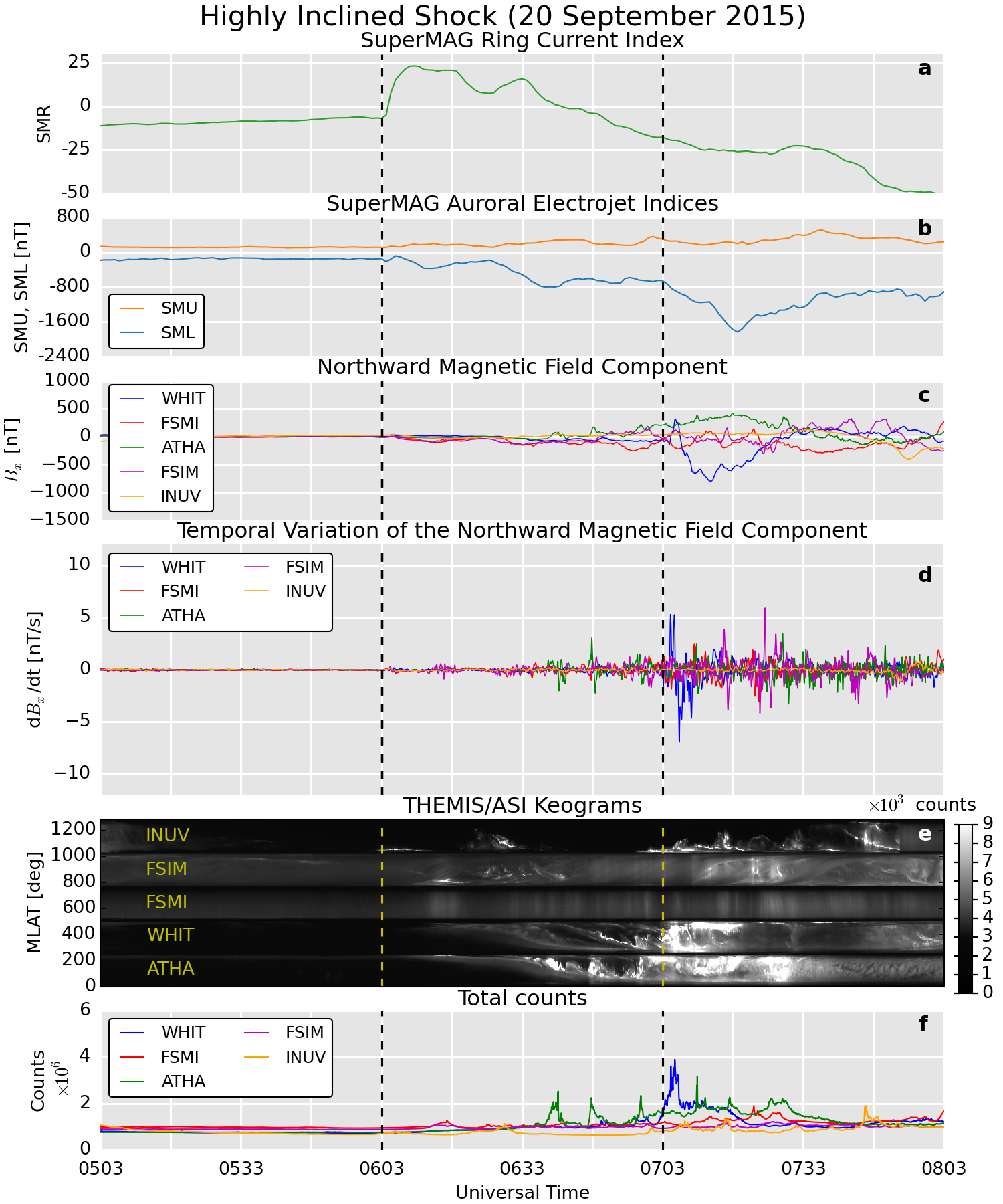}
			\caption{Summary plots of ground magnetometers (SuperMAG and THEMIS, a-d), and THEMIS/ASI (e and f) data for the highly inclined shock of 20 September 2015.}
			\label{522}
		\end{figure*}

		Panel c shows $B_x$ values recorded by McGrath (MCGR), Gakona (GAKO), Inuvik (INUV), Fort Yukon (FYKN), and White Horse (WHIT). Geographic coordinates and geomagnetic conditions of these stations around substorm onset are summarized in the top part of Table \ref{table1}. The stations' positions with respect to the 5 April 2010 midnight MLT (magenta line) are shown in Figure \ref{stations}. These stations were selected for two reasons: first, these stations have available THEMIS/ASI data with minimal data gaps, and, second, they were located in the northwest region of North America during the occurrence of the April 2010 substorm. The panel shows moderate field intensifications recorded by all stations after shock impact and before substorm onset most likely due to the propagation of magnetospheric Alfv\'en waves triggered by the shock impact \cite{Carter2015,Oliveira2018b}. Consequently, panel d shows very intense magnetic \dbxdt{} variations immediately after the substorm onset, particularly in the cases of MCGR (8 nT/s), FYKN (--9 nT/s), and GAKO (--22.6 nT/s). Intense and moderate \dbxdt{} variations are still seen until 1011 UT, more than 1 hour after substorm onset. \par

		Figure \ref{394}e shows some white-light auroral brightening between 0826 and 0903 UT, most likely due to the propagation of the Alfv\'en waves mentioned above, particularly at FYKN. However, a clear substorm onset at 0903 UT (second vertical yellow dashed line) is marked by a sudden poleward expansion of the auroral oval, as clearly seen from $\sim$66.7$^\circ$ MLAT (WHIT) to $\sim$71.1$^\circ$ MLAT (INUV). Similar observations during intense magnetic storms were reported by \citeA{Ngwira2018c} with THEMIS/ASI data. Although such auroral oval intensification and poleward expansion are known common features of substorms \cite<e.g.,>[]{Akasofu1964a,Zhou2001,Meurant2004}, here we explore for the first time the effects of shock impact angles on such auroral oval response during isolated substorms. \par

		\begin{figure}[t]
			\centering
			\includegraphics[width = 12cm]{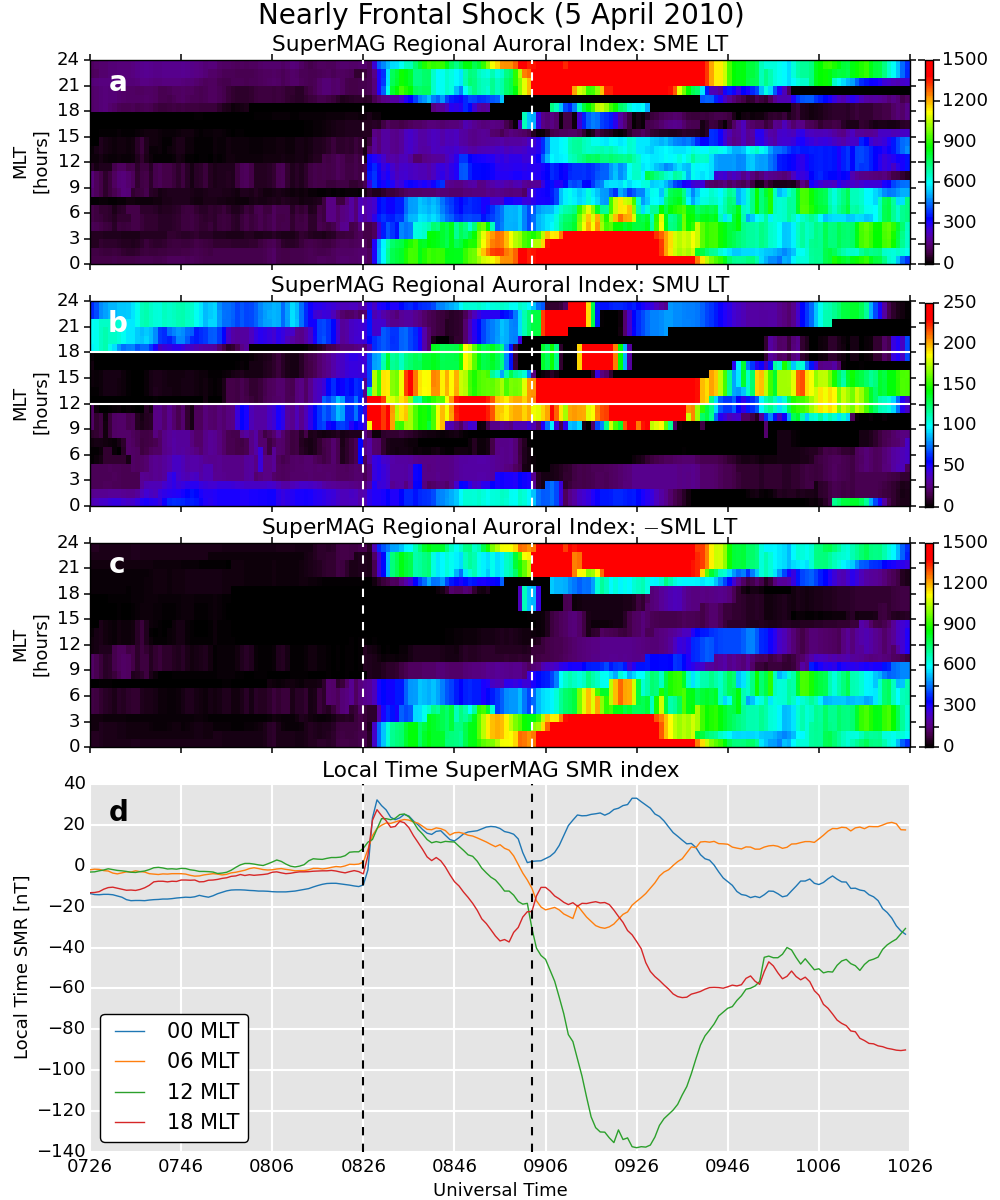}
			\caption{Regional SuperMAG indices for the nearly frontal shock of 5 April 2010: panels (a), (b), and (c) show the 0-23 magnetic local time SME, SMU, and $-$SML indices; and (d) shows the SMR00, SMR06, SMR12, and SMR18 indices for the midnight, dawn, noon and dusk sectors, respectively.}
			\label{SuperMAG_394}
		\end{figure}

		Figure \ref{522} is similar to Figure \ref{394}, but it is for the HIS (20 September 2015). An SI$^+$ signature at 0603 UT is evident in panel a, with a rise time of $\sim$5 min and amplitude $\sim$30 nT. This rise time is longer than the rise time caused by the NFS case because the HIS compressed the magnetosphere asymmetrically leading to a slow response of the magnetospheric currents closest to Earth \cite{Guo2005,Wang2006a,Rudd2019}, and the similarity of the SI$^+$ amplitudes in both events indicate that both shocks were indeed quite strong. A substorm onset is identified at 0703 UT, 1 hour after HIS impact, as characterized by a moderately sharp and gradual decrease in the SML index from --750 nT to --1800 nT. Such relatively weaker and slower response is a well-known signature of substorms triggered by IP shocks with large shock impact angles \cite{Oliveira2014b,Oliveira2015a}. \par

		Moderate magnetic field responses are seen at WHIT, Fort Smith (FSMI), Athabaska (ATHA), Fort Simpson (FSIM), and INUV, with WHIT being the strongest. The bottom part of Table \ref{table2} shows geographic coordinates and geomagnetic conditions at substorm onset for these stations. The black line in Figure \ref{stations} corresponds to MLT = 24 hr on 20 September 2015 at 0703 UT. Figure \ref{522}d shows \dbxdt{} variations caused by wave activity. Panel d also shows weak \dbxdt{} activity following the substorm onset, except for WHIT with two very clear peaks 2-3 minutes after substorm onset (5 nT/s), and another peak 5 minutes after substorm onset (--7 nT/s). Another peak surpassing 5 nT/s is observed at FSIM $\sim$20 minutes after substorm onset. Some weak \dbxdt{} activity is observed later but \dbxdt{} never surpasses 5 nT/s. \par

		Auroral brightening during the substorm triggered by HIS is shown in Figure \ref{522}e. These observations are clearly weaker and occur later in comparison to the auroral brightening triggered by the NFS. In addition, a poleward expansion of the auroral oval as seen in Figure \ref{394}e and reported by \citeA{Ngwira2018c} is not clearly observed in response to the substorm activity triggered by the HIS. \par

		\begin{figure}[t]
			\centering
			\includegraphics[width = 12cm]{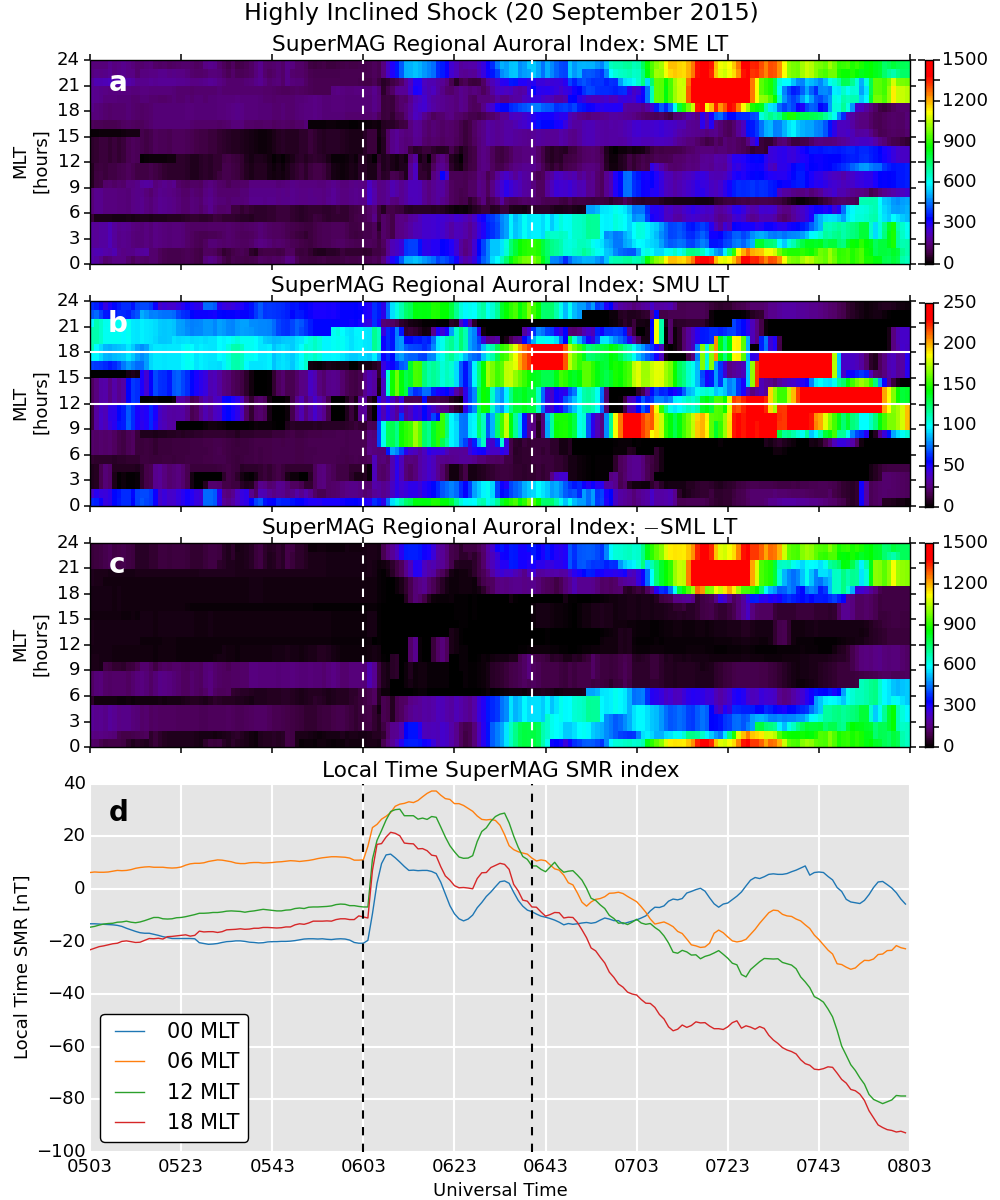}
			\caption{Regional SuperMAG indices for the highly inclined shock of 20 September 2015: panels (a), (b), and (c) show the 0-23 magnetic local time SME, SMU, and $-$SML indices; and (d) shows the SMR00, SMR06, SMR12, and SMR18 indices for the midnight, dawn, noon and dusk sectors, respectively.}
			\label{SuperMAG_522}
		\end{figure}

		The events evaluated in this study fall shortly before the very intense substorm classifications suggested by \citeA{Tsurutani2021}: NFS, almost a supersubstorm, and HIS, almost an intense substorm. As suggested by a reviewer, we looked into possible effects the shock impact angles can cause on the regional SuperMAG auroral and ring current indices. \par

		Figures \ref{SuperMAG_394} and \ref{SuperMAG_522} show regional SuperMAG auroral indices for 00-23 MLT (panels a, b, c) and the local time SMR indices for 00, 06, 12, and 18 MLT (panel d) for both shocks, respectively. The nearly head-on impact causes the SMR indices to increase nearly simultaneously at all local time sectors. Conversely, the regional SMR indices respond in a different way to the HIS: the rise time of SMR at 18 MLT is shorter than the rise time of SMR at 06 MLT because the shock impacted the magnetosphere in the afternoon sector, confirming the features of both shock impact angles \cite{Wang2006a,Samsonov2015,Rudd2019}. The SME and SML indices are very symmetric for the NFS, peaking around 24 MLT, while the same indices are quite asymmetric in the case of the HIS, being more intense in the pre-midnight region. On the other hand, the regional SMU index is highly enhanced at the NFS-triggered substorm onset in the afternoon sector with some enhancements near the dusk sector. The SMU enhancement on 5 April 2010 occurs while SMR at 18 MLT increases, indicating a strong effect on the partial ring current predominantly located near the dusk sector These observations agree with the results reported by \citeA{Fu2021}, who showed that an additional current wedge is highly intensified in the dusk sector moving toward the noon sector during very intense substorms. However, this feature is not quite well captured during the substorm triggered by the NFS because the dusk sector at 0903 UT on 5 April 2010 corresponded to the northern Russian territory, which is notably known for its low number of ground magnetic stations (see Figures S16 and S17 in the supporting information). On the other hand, the SMU response after the HIS onset reaches areas closer to the dusk sector, but such intensifications are clearly weaker. These HIS results are also in agreement with the results provided by \citeA{Fu2021}, and we believe that the shock impact angle affects the dynamic of the partial ring current by its effect on the subsequent triggering of intense substorms and even supersubstorms. \par

	\subsection{Magnetic field and plasma observations in geospace}

\begin{figure}[t]
			\centering
			\includegraphics[width=12cm]{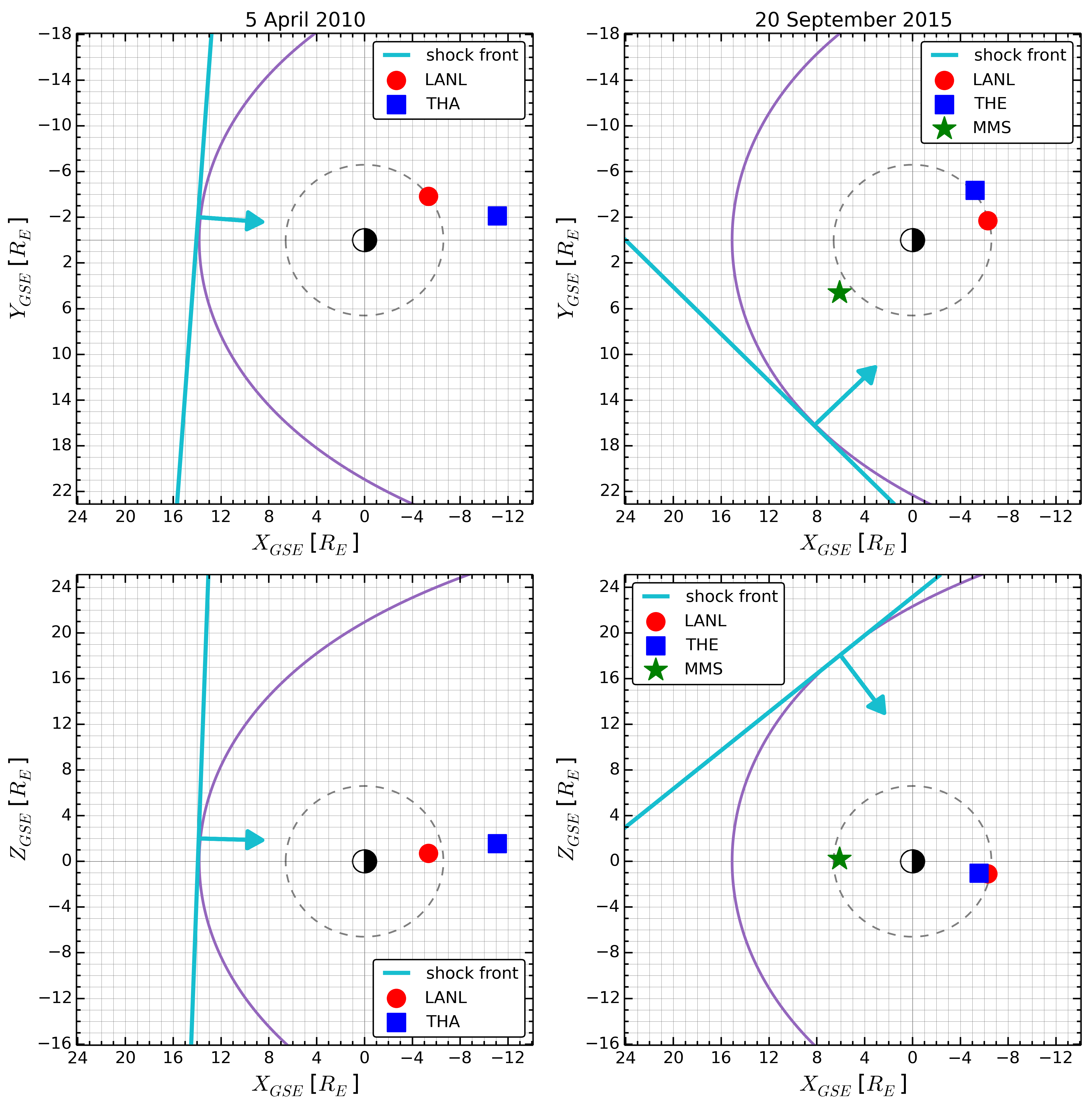}
			\caption{Cartesian geographic positions of the LANL and THEMIS satellites in geospace at the onsets of substorms triggered by the nearly frontal shock on 5 April 2010 at 0903 UT (first column) and the highly inclined shock on 20 September 2015 at 0703 UT (right column). Positions are plotted in the xy plane (top row) and in the xz plane (bottom row). The solid cyan lines and cyan arrows indicate the shock front and shock normal orientations in the respective plane. The solid purple curves show the magnetopause produced by the \citeA{Shue1998} empirical model, whereas the dashed black circles indicate the geosynchronous orbit. The green star on the right column indicates MMS positions in the equatorial and meridional planes; MMS data are only available for the second shock event.}
			\label{orbits}
		\end{figure}

		Figure \ref{orbits} shows the orbit positions of satellites in GSE coordinates plotted in the equatorial plane (upper row) and meridional plane (lower row) for the NFS (left column) and HIS (right column). The left column panels show the THEMIS-A (THA, blue square) and LANL-1989-046 (LANL, red circle) positions at the NFS impact time, whereas the right column panels show THEMIS-E (THE), LANL-01A (LANL), and MMS (green star) positions at the HIS impact time. The magnetopause positions (purple lines) are plotted with the \citeA{Shue1998} empirical model. The thick cyan lines represent the shock front right before impact and the cyan arrows represent the shock normal orientation in the respective plane. The dashed black circles represent the geosynchronous orbit. \par

		\begin{figure*}
			\centering
			\includegraphics[width=15cm]{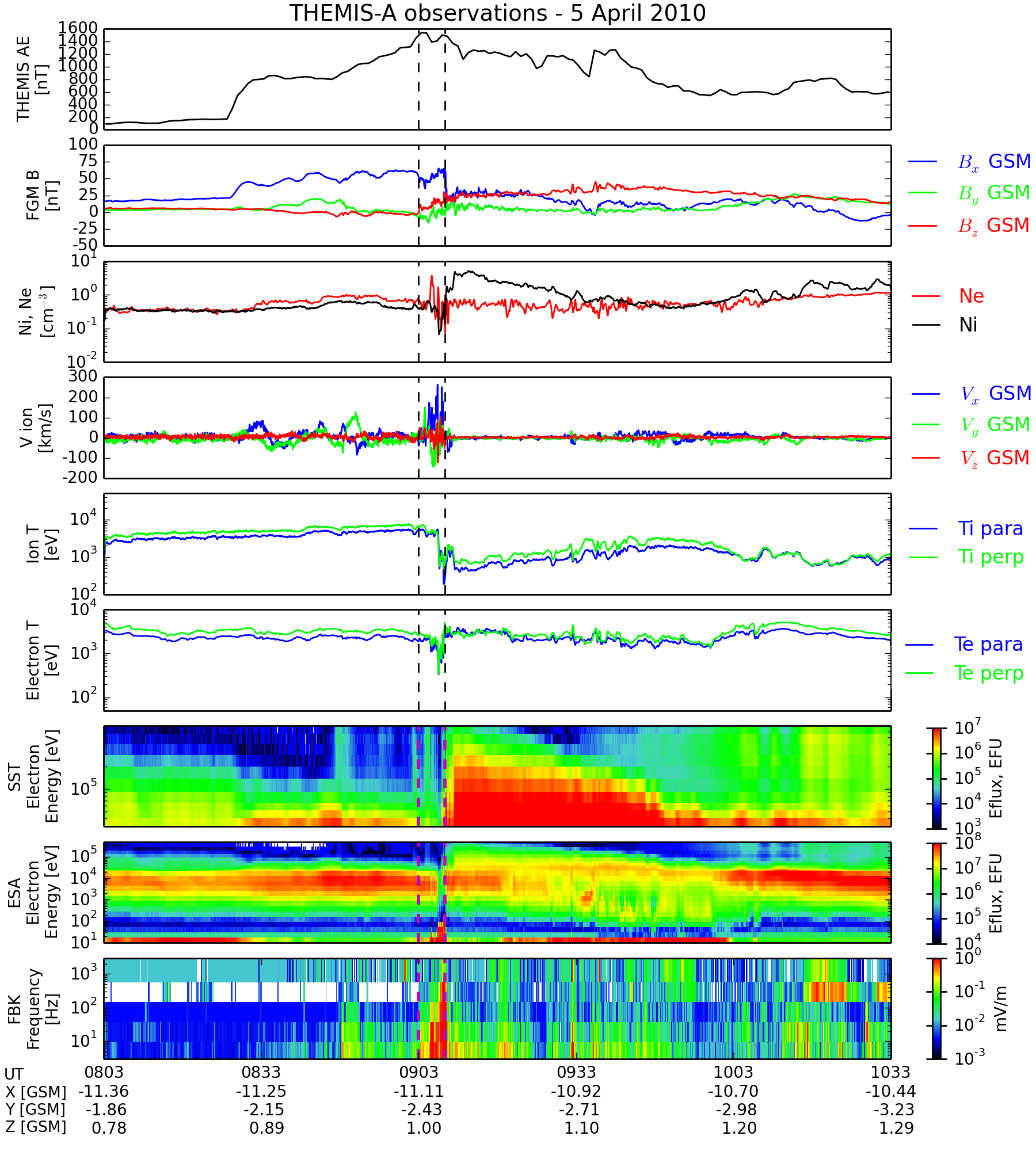}
			\caption{THEMIS observations during a period of substorm activity triggered by the nearly frontal shock on 5 April 2010. From top to bottom: THEMIS AE index; magnetic field, electron and ion densities; plasma velocity, ion.electron parallel and perpendicular temperatures; EFU electron flux; SST electron flux; and electric field.}
			\label{tha}
		\end{figure*}

		\begin{figure*}
			\centering
			\includegraphics[width=15cm]{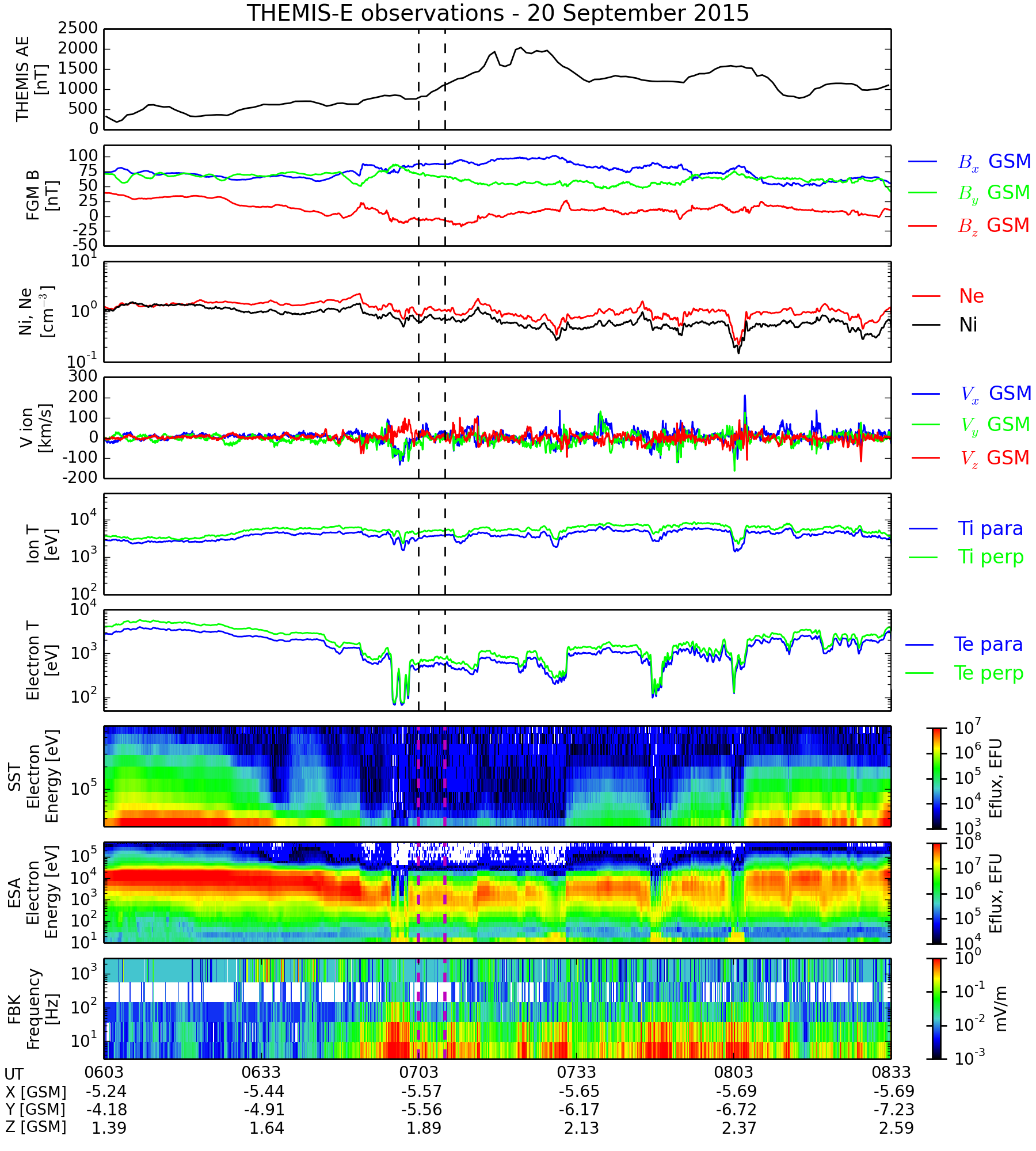}
			\caption{THEMIS observations during a period of substorm activity triggered by the highly inclined shock of 20 September 2015. From top to bottom: THEMIS AE index; magnetic field, electron and ion densities; plasma velocity, ion.electron parallel and perpendicular temperatures; EFU electron flux; SST electron flux; and electric field.}
			\label{the}
		\end{figure*}

		Magnetic field and plasma response during the 5 April 2010 substorm is shown in Figure \ref{tha}. From top to bottom, the figure shows: the THEMIS AE index; three components of the FGM magnetic field in Geocentric Solar Magnetospheric (GSM) coordinates; electron and ion densities; three components of the ion velocity; electron and ion parallel and perpendicular temperatures; SST and ESA electron fluxes; and filter bank (FBK) electric field. \par

		Figure \ref{tha} shows observations recorded by THA during the NFS-triggered substorm. THA was at (--11.1, --2.4, 0.9) \RE{} at 0903 UT, meaning the spacecraft was slightly past midnight down the tail (Figure \ref{orbits}). The \citeA{Tsyganenko2005} model indicates that the THA magnetic footprints were over the Canadian Northwest Territories (white square with a magenta frame in Figure \ref{stations}). There is a sharp increase in the THEMIS AE index at 0826 UT caused by the NFS impact, and a peak in this index is observed shortly before substorm onset. There are moderate and sharp decreases in $B_x$ and $B_y$, and a moderate increase in $B_z$ at 0903 UT. However, within the following 6 minutes, ion/electron parallel and perpendicular temperatures decrease and then sharply increase at 0906 UT, with electron and ion densities severely fluctuating in this 6-minute interval. The differences in ion and electron densities are most likely associated with contaminations caused by photoelectrons with energies $<$ 10 eV \cite{McFadden2008}. The very intense ion velocity component $V_x$ reaches positive values nearly 300 km/s, which indicates plasma moving Earthward during the substorm-like energetic particle injection. The electron fluxes measured by the SST and ESA instruments also increase at 0909 UT, and the electric field observed by the FBK instrument increases at 0903 UT and peaks at 0909 UT. Several intense peaks with $|$\dbxdt$|$ $>$ 5 nT/s were recorded on the ground by MCGR, GAKO, INUV, and FYKN in North America after 0909 UT (Figure \ref{394}d), being well timed with the energetic particle injection observed by THA. \par 

		Magnetic field and plasma measurements during the HIS-triggered substorm are provided by THE (Figure \ref{the}). THE was positioned at (--5.6, --5.6, 1.9) \RE{} at 0703 UT, therefore the satellite was at midway between midnight and dawn positions (Figure \ref{orbits}). THE's footprints were in a region over the border of Manitoba and Ontario (white square with a black frame in Figure \ref{stations}). Although the THEMIS AE index starts to increase gradually at substorm onset (0703 UT) and THE observes very weak increases in the electron parallel and perpendicular temperatures, THE observes a few minutes after 0703 UT a weak earthward ion flow with $V_x$ $<$ 50 km/s along with very weak enhancements in electron fluxes and electric field. This modest energetic particle injection could be related to modest field variations on the ground ($|$\dbxdt$|$ $<$ 1.5 nT) seen for all stations in Figure \ref{522}d, except for WHIT ($|$\dbxdt$|$ slightly larger than 5 nT/s). \par

		\begin{figure}[t]
			\centering
			\includegraphics[width=12cm]{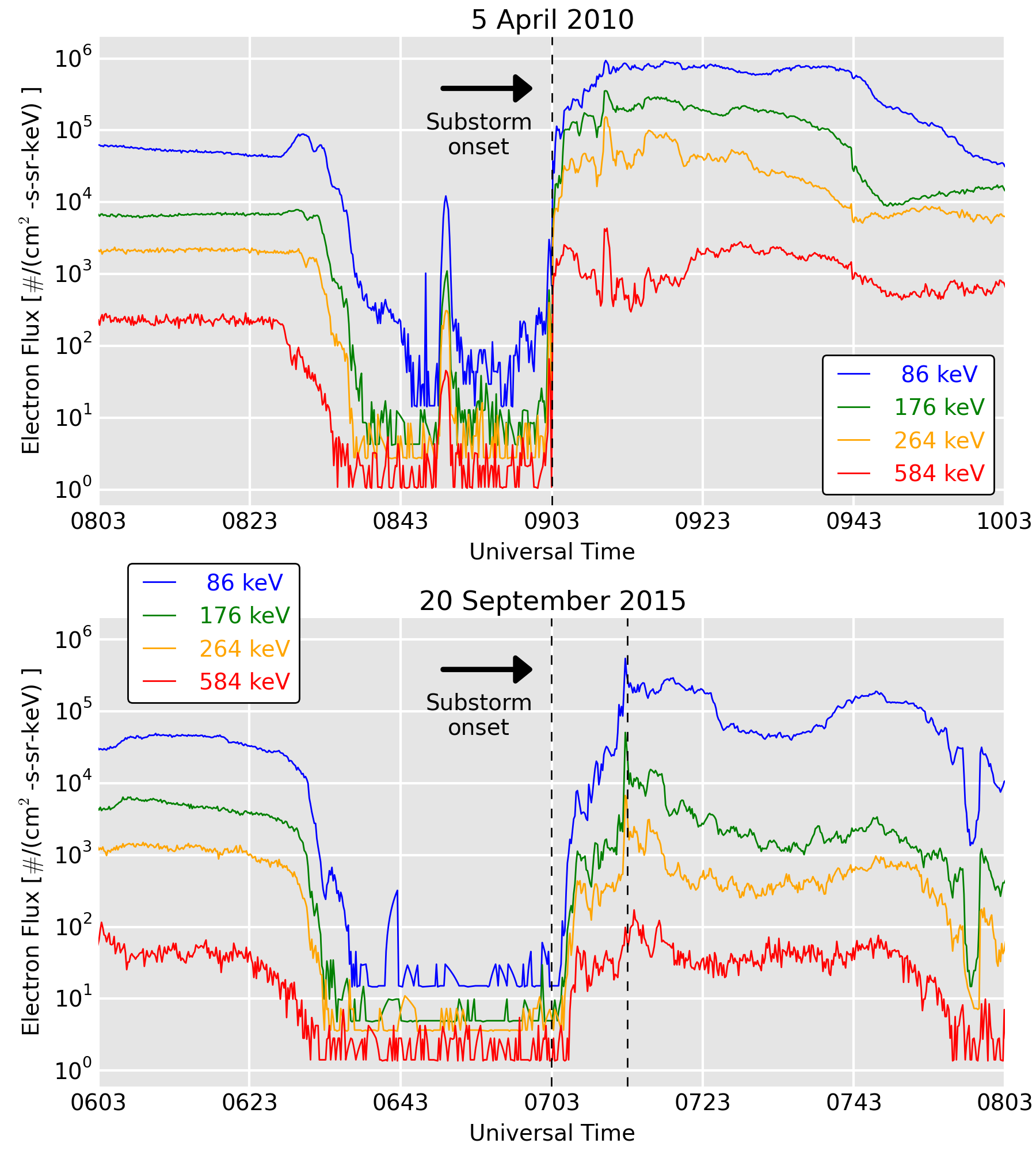}
			\caption{Electron fluxes measured in geosynchronous orbit by LANL-1989-046 on 5 April 2010 (upper panel) and by LANL-01A on 20 September 2015 (lower panel). Four different energy channels are used: 86 keV (blue); 176 keV (green); 264 keV (yellow); and 584 keV (red). The substorm onsets occurred at 0903 UT and 0703 UT, respectively (vertical dashed back lines at midpoint), and were followed by intense energetic particle injections. The second vertical dashed black line in the bottom panel indicates a secondary energetic particle injection observed by LANL around 0713 UT on 20 September 2015.}
			\label{lanl}
		\end{figure}

		LANL data are now shown to supplement the observations performed by THEMIS. LANL/SOPA observations of electron flux near geosynchronous orbit are shown in Figure \ref{lanl} for the NFS (upper panel) and HIS (lower panel). The plots are centered at $\pm$1 hr around the respective substorm onset and show electron flux in the following energy channels: 86 keV (blue), 176 keV (green), 264 keV (yellow), and 584 keV (red). Both LANL satellites were in the post-midnight regions during the onsets of both substorms (Figure \ref{orbits}). \par

		Electron fluxes before substorm onsets in both cases are very similar, ranging from around $\sim10^2-10^5$ electrons/(cm$^2$ $\cdot$ s $\cdot$ sr $\cdot$ keV). In both cases, there are two growth phase flux dropouts at $\sim$0833 and $\sim$0846 UT for the NFS case, and $\sim$0626 and $\sim$0642 UT for the HIS case. It is known that such growth phase dropouts are caused by the stretching of the geomagnetic field toward the tail near local midnight, and such dropouts often precede nightside magnetospheric substorm-like energetic particle injections \cite{Sauvaud1992,Reeves2003}. In addition, as shown by the bottom panels of Figures \ref{tha} and \ref{the}, the overall electron fluxes are smaller after substorm onsets due to electron acceleration in the plasmasheet, as observed with THEMIS data by \citeA{Sivadas2017}. There is a sharp increase in all energy levels at the NFS-triggered substorm (0903 UT), indicating a rapid and symmetric compression of the magnetotail. This energetic particle injection coincides with intense \dbdt{} peaks observed particularly by GAKO, MCGR, and FYKN, with the GAKO \dbdt{} peak surpassing --22 nT/s. Conversely, the energetic particle injection in the second case occurs $\sim$2 minutes (0705 UT) after the HIS-triggered substorm, and another sharp injection is seen at 0713 UT (second vertical dashed black line in the bottom panel of Figure \ref{lanl}). Both particle injections coincide with intense \dbdt{} variations observed by WHIT (Figure \ref{522}d). These energetic particle injections observed by LANL in September 2015 are more complex and more asymmetric than the energetic particle injections observed by LANL in April 2010, with the former most likely being due to the asymmetric magnetotail compression by the inclined shock. 

		\begin{figure*}
			\centering
			\includegraphics[width=14cm]{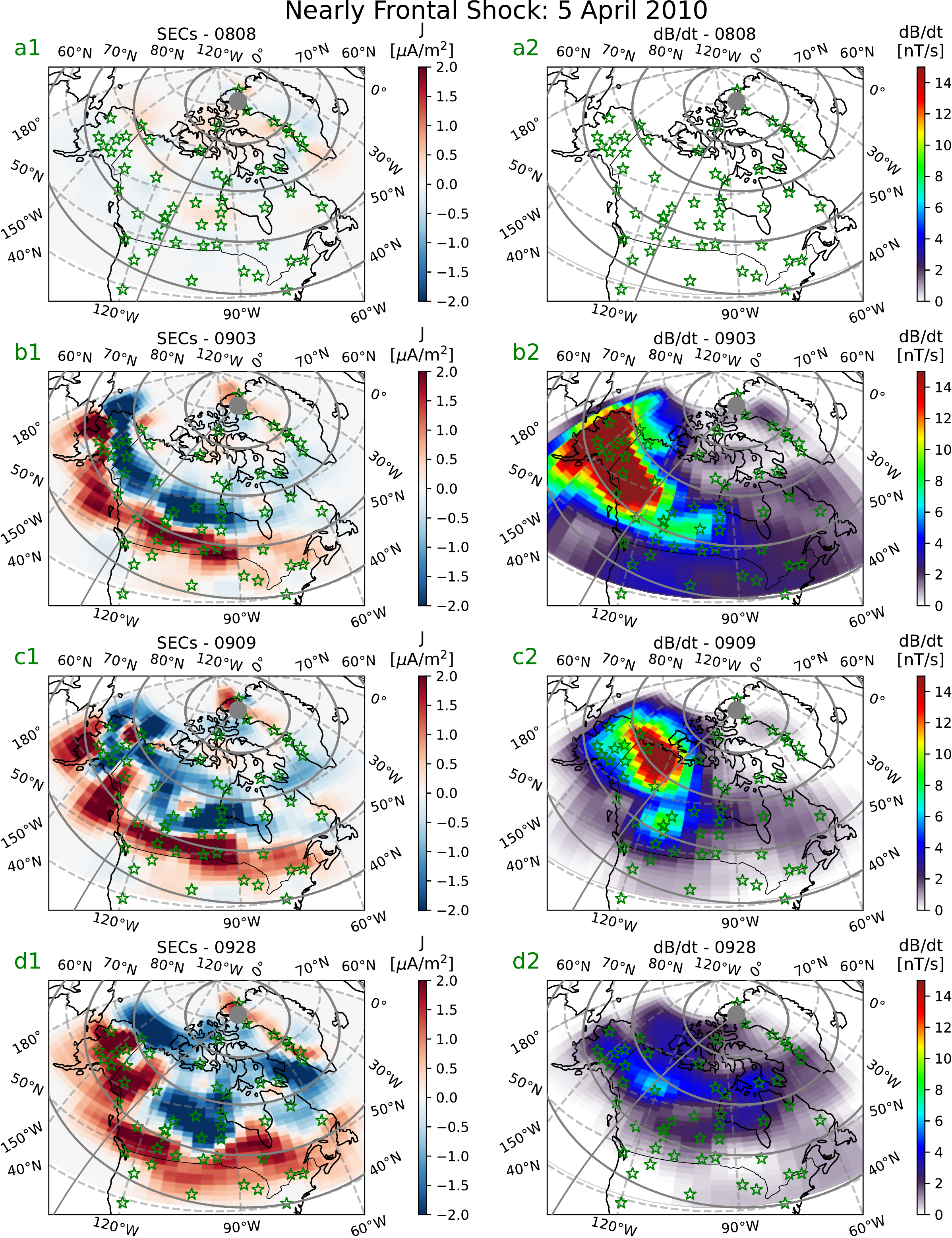}
			\caption{Ionospheric currents (left column) and ground \dbdt{} variations (right column) computed with the SECS method for the 5 April 2010 interplanetary shock. In the left column, blue colors indicate downward Region 1 currents, whereas red colors indicate upward Region 2 currents. The green stars indicate the geographic locations of the stations used to compute the ionospheric currents and ground \dbdt{} variations shown in the figure. More information about these stations can be found in the supporting information.}
			\label{sec_dbdt1}
		\end{figure*}

		\begin{figure*}
			\centering
			\includegraphics[width=14cm]{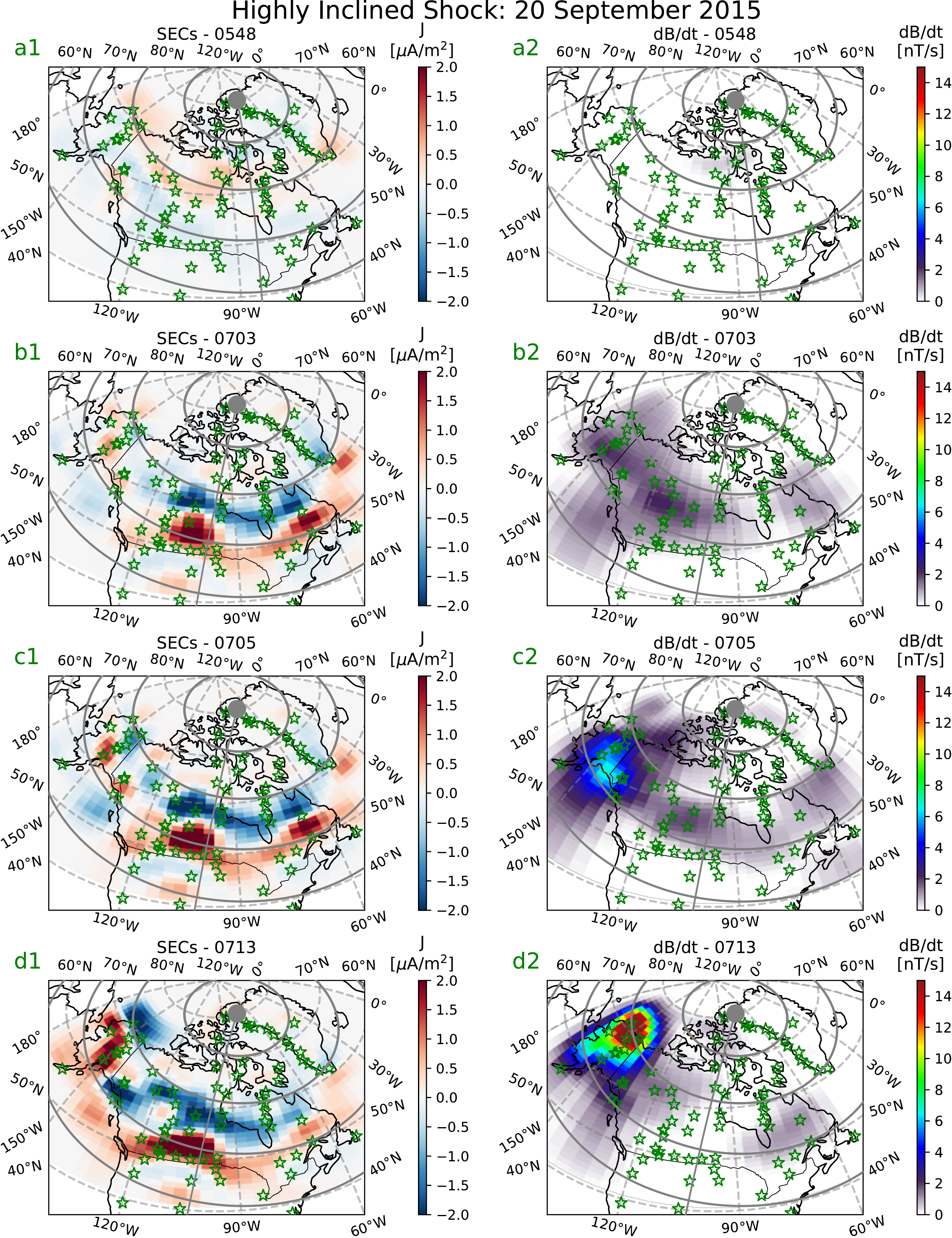}
			\caption{Ionospheric currents (left column) and ground \dbdt{} variations (right column) computed with the SECS method for the 20 September 2015 interplanetary shock. In the left column, blue colors indicate downward Region 1 currents, whereas red colors indicate upward Region 2 currents. The green stars indicate the geographic locations of the stations used to compute the ionospheric currents and ground \dbdt{} variations shown in the figure. More information about these stations can be found in the supporting information.}
			\label{sec_dbdt2}
		\end{figure*}

	\subsection{Ionospheric current and \dbdt{} response produced by the SECS method}

		In this section we evaluate the response of the auroral ionospheric currents over North America and Greenland and the subsequent ground \dbdt{} variations using the spherical elementary current system (SECS) technique developed by \citeA{Amm1999}. SECS uses inverted fluctuations of ground magnetometer data in a decomposition method of singular values to compute ionospheric currents \cite{Weygand2011}. There are two outputs from the SECS inversion. The first output is the equivalent ionospheric currents, which are a combination of the real Hall and Pedersen currents. One of the important features of this technique is that it requires no integration time of the magnetometer data. The temporal and spatial resolutions of the equivalent ionospheric currents are 10 s and 6.9$^\circ$ geographic longitude (GLON) by 2.9$^\circ$ geographic latitude (GLAT). The second output is the SEC amplitudes, which are a proxy for the field-aligned-like currents, with a 10 s temporal resolution and 3.5$^\circ$ GLON by 1.5$^\circ$ GLAT spatial resolution. Both sets of currents are derived for an altitude of 100 km. The supporting information provides further information on the ground magnetometer stations located in North America and Greenland whose data were used in the SECS procedure. \par

		\begin{figure}[t]
			\centering
			\includegraphics[width = 10cm]{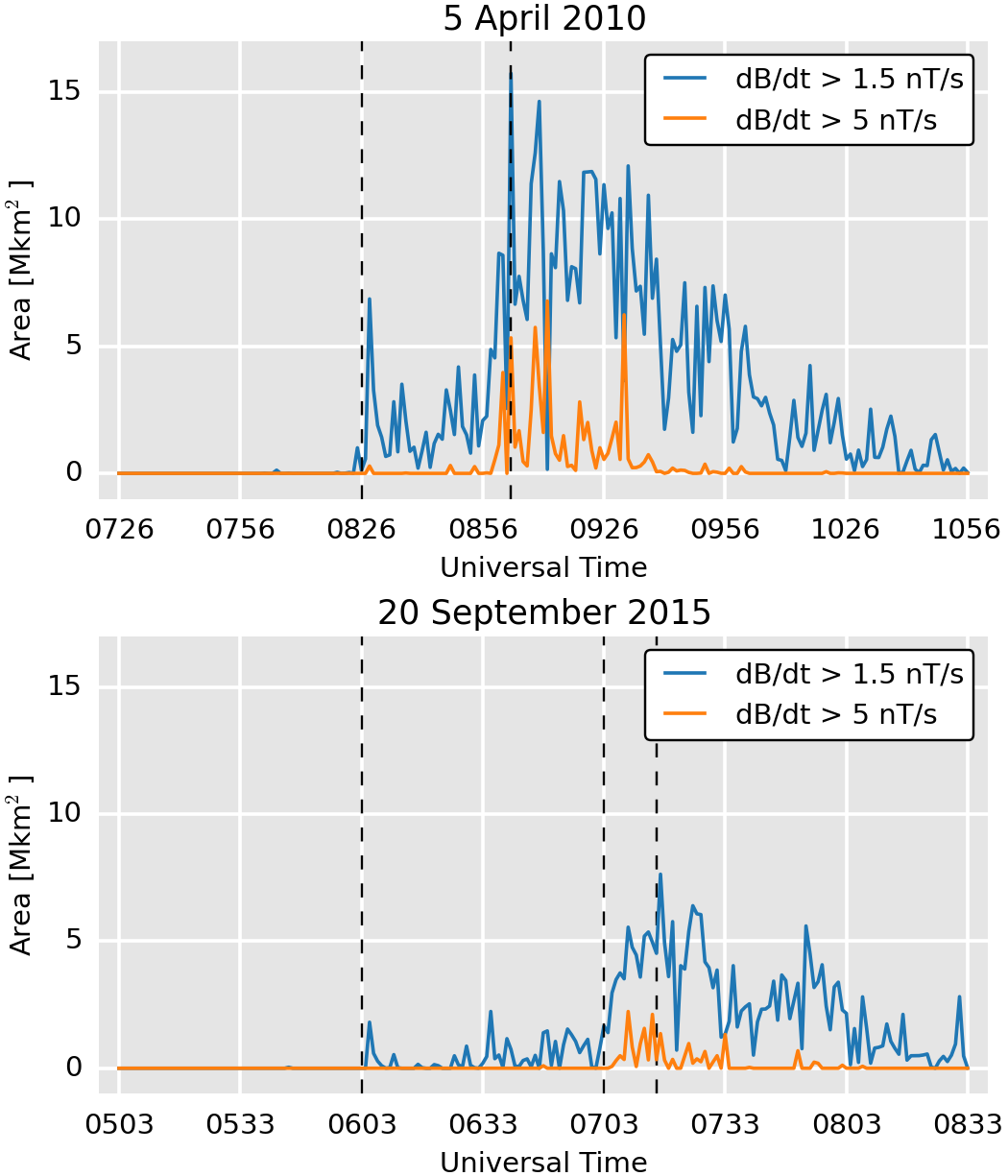}
			\caption{Top panel: time evolution of regions with geographic areas where dB/dt surpassed the threshold of 1.5 nT/s (blue line) and 5 nT/s (orange line) during the substorm activity caused by the nearly frontal shock. Bottom panel: the same, but for the highly inclined shock. In both panels, the first vertical dashed black lines indicate the respective shock impacts, whereas the second vertical dashed black lines indicate the respective substorm onsets. The third vertical dashed black line, at 0713 UT, in the bottom panel indicates the time of maximum areas with \dbdt{} $>$ 1.5 nT/s, which coincides with the second energetic particle injection observed by LANL, and a \dbdt{} peak observed by WHIT.}
			\label{areas}
		\end{figure}

		Figure \ref{sec_dbdt1} shows the SECS results for the NFS. The left column is for the ionospheric SECs, while the right column is for \dbdt{} = $\sqrt{(dB_x/dt)^2 + (dB_y/dt)^2}$, with \dbydt{} being the temporal variation of the eastward component of the geomagnetic field $B_y$. The cadence of SECS data here is 1 minute. The first row shows SECS results during quiet times before shock impact (0808 UT), the second row at substorm onset (0903 UT), and the subsequent rows later at 0909 and 0928 UT. The supporting information provides a movie with the temporal evolution of the ionospheric currents and \dbdt{} variations produced by the SECS method with the intervals shown in Figure \ref{394} and Figure \ref{522} for both shocks, respectively. The thick grey lines correspond to MLATs from 50$^\circ$ toward the pole in 10$^\circ$ increments. The grey ``clock-hand-like" lines correspond to MLT = 24 hr. The green stars indicate the positions of the stations whose data were used in the SECS procedure (see supporting information). \par

		The left panel in the second row of Figure \ref{sec_dbdt1} shows clear Region 1 currents below MLAT = 70$^\circ$ (downward, blue) and Region 2 currents below MLAT = 60$^\circ$ (upward, red) \cite{Iijima1976a,Cowley2000} over North America. On the other hand, the right panel in the second row shows that \dbdt{} surpassed 1.5 nT/s in almost all North America and northwestern Greenland, with \dbdt{} $>$ 5 nT/s in almost all Alaskan areas and most of the northwestern Canada. These intense \dbdt{} variations occur shortly after the substorm onset triggered by the NFS and coincide with a sharp enhancement in electron flux observed by LANL (Figure \ref{lanl}, top panel). The third row shows that \dbdt{} variations are still strong at 0909 UT (right panel) but they are weaker than the \dbdt{} variations at substorm onset and coincide with intense energetic particle injections observed by THA at 0909 UT (Figure \ref{tha}). The same occurs with the ionospheric SECs (left panel), but the \citeA{Iijima1976a} current pattern is not as clear as it was at substorm onset. Although ionospheric currents are still strong at 0928 UT (bottom row, left panel, but weaker in comparison to the currents at 0903 UT, \dbdt{} variations are relatively weaker probably due to low level of ionospheric current variations, but \dbdt{} variations can yet surpass 5 nT/s in most of Alaska and northern Canada. \par

		Similar SECS data are plotted for the substorm triggered by the HIS on 20 September 2015 in Figure \ref{sec_dbdt2}. The two panels in the second row show very mild ionospheric and ground \dbdt{} responses at substorm onset (0703 UT), with relatively weaker Region 1 and Region 2 currents over central Canada, and some regions in Alaska and Northwestern Territories with \dbdt{} surpassing the 1.5 nT/s threshold. These current and \dbdt{} intensifications coincide with a weak energetic particle injection observed by THE at substorm onset (Figure \ref{the}). Two minutes later, the left panel in the third row shows slightly stronger current values and a significant region over southwestern Alaska with \dbdt{} $>$ 6 nT/s. These SECS results coincide with the first peak in energetic particle injection observed by LANL at 0705 UT (Figure \ref{lanl}, bottom panel). However, 8 minutes later, a more clear current system appears over North America (bottom row, left panel), whereas \dbdt{} variations larger than 4 nT/s are seen over all Alaska, and extremely high \dbdt{} variations ($>$ 10 nT/s) is seen in northern Alaska. These results are the peak of the SECS results, and coincide with the second energetic particle injection observed by LANL at 0713 UT. \par

		Figure \ref{areas} shows the areas in Mkm$^2$ with \dbdt{} thresholds of 1.5 nT/s (blue line) and 5 nT/s (orange line) caused by the IP shock impacts. These area values are extracted from the maps with \dbdt{} variations computed with the SECS technique. The top panel is for the NFS, whereas the bottom panel is for the HIS. The first vertical black lines corresponds to the respective shock impact (0826 and 0603 UT), whereas the second vertical black lines (0903 and 0703 UT) correspond to the respective substorm onset. Shortly after shock impacts areas with \dbdt{} $>$ 1.5 nT/s peak at 17 Mkm$^2$, and the largest area value with \dbdt{} $>$ 5 nT/s occurs at 0911 UT (7 Mkm$^2$). The results shown in the bottom panel for the HIS are quite different. The general areas peak at smaller values compared to the NFS case, with \dbdt{} $>$ 1.5 nT/s peaking at 7.5 Mkm$^2$ (0713 UT, third vertical line in the bottom panel), and areas with \dbdt{} $>$ 5 nT/s peaking at 2.5 Mkm$^2$ (0708 UT). The area peak at 0713 UT coincides with the secondary energetic particle injection peak observed by LANL, and a \dbdt{} peak observed by WHIT on the ground. Therefore, the SECS results for the HIS are weaker and occur later in comparison to the NFS case, as expected when geoeffectiveness of nearly frontal and highly inclined shock impacts are compared \cite{Oliveira2018a}. Additionally, these peaks in geographical areas surpassing the \dbdt{} thresholds coincide with the substorm-like energetic particle injections observed by THEMIS and LANL discussed above. \par

	\subsection{Magnetospheric ULF wave contributions to ground \dbdt{} variations}

		ULF waves are known to affect magnetospheric dynamics immediately after shock impacts and during magnetic storms, including drift-bounce resonance with charged particles to form storm-time energetic particles \cite<e.g., >[and references therein]{Menk2011,Zong2012}. Additionally, \citeA{Kim2010} observed Pi2 pulsations (frequency range 7-25 mHz) associated with flow bursts in the magnetotail, \citeA{Baumjohann1984} observed Pi2 pulsations during substorms, and \citeA{Ngwira2018c} observed intense Pc5 (frequency range 1.5-5 mHz) wave activity around the dawn and midnight sectors associated with intense \dbdt{} variations in these regions. Pc5 pulsations are known to occur during GIC-related events \cite{Pulkkinen2005,Heyns2020,Yagova2021}. Therefore, it is expected to observe strong ULF wave activity in the events analyzed in this study.  \par

\begin{figure}[t]
			\centering
			\includegraphics[width = 15cm]{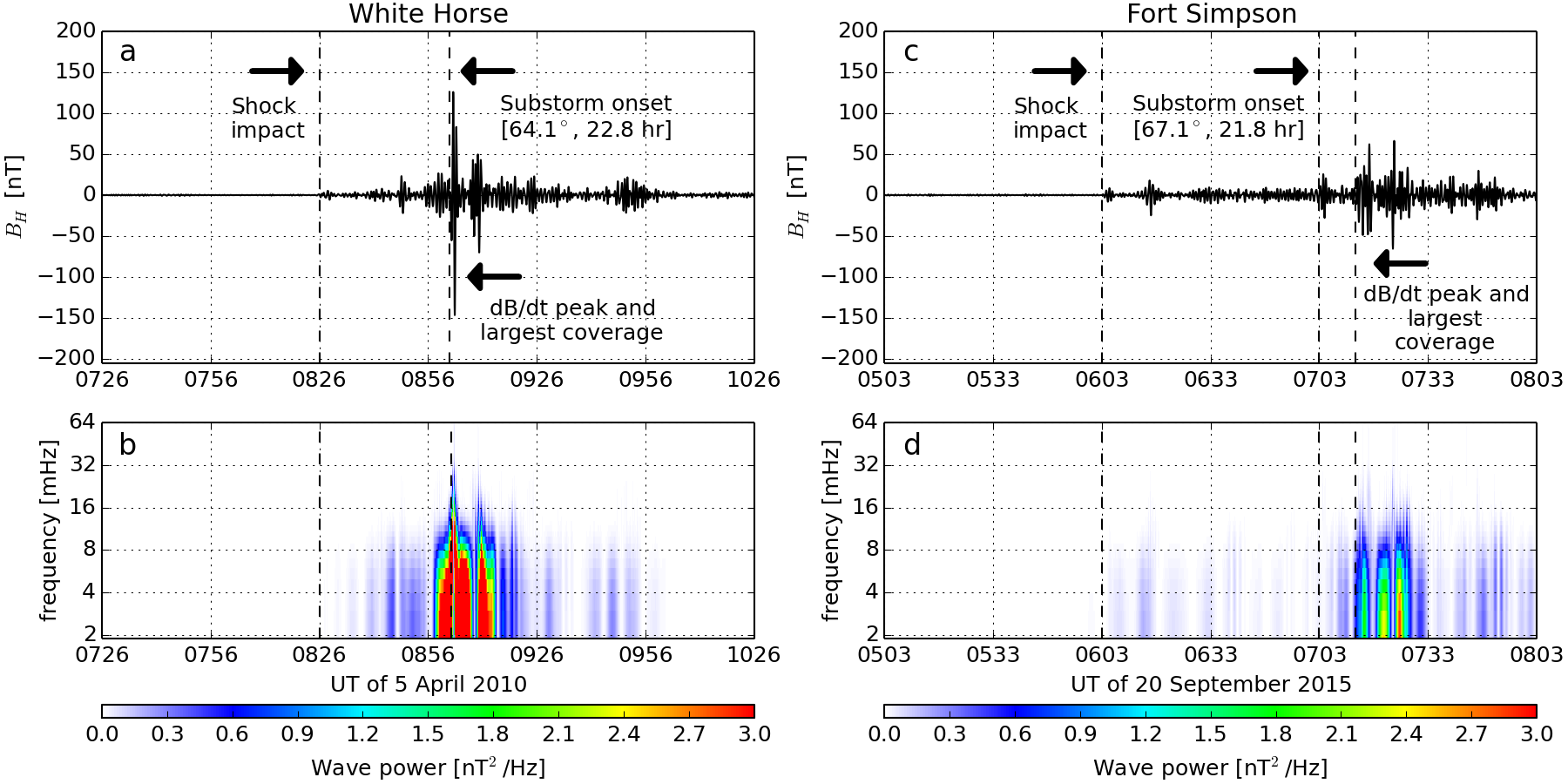}
			\caption{ULF wave analyses with WHIT data for the nearly frontal shock (left column) and with FSIM data for the highly inclined shock (right column). The upper row shows Pc5 waves with frequencies filtered in the range 2-4 mHz, while the lower row shows waves in a broader range of frequencies (2-64 mHz) with wave power being color coded. WHIT and FSIM were selected because they occupied similar MLAT and MLT positions at the respective substorm onset.}
			\label{waves}
		\end{figure}

		We then evaluate ULF wave response in Figure \ref{waves} with WHIT data for the NFS (left column), and with FSIM data for the HIS (right column). The MLATs and MLTs of both stations near substorm onset are very similar and they are shown in the respective panel in the top row. The first row shows Pc5 waves with frequencies filtered in the range 2-4 mHz, whereas the second row shows wave power plotted as a function of UT and frequency for ULF waves in a broader frequency range (2-64 mHz). We used the band-pass filtered (2-64 mHz) measurements to obtain wave power via the fast Fourier transformation method \cite{Frigo1998}. \par

		Pc5 pulsation activity is seen after both shock impacts (first dashed vertical lines in panels a and c). Wave activity begins 7 minutes before substorm onset on 5 April 2010 (0856 UT), but a peak of 130 nT is seen at substorm onset (0903 UT). The same trend is seen in wave power peaking near 32 mHz, but intense wave power levels ($>$ 5 nT$^2$/Hz) are seen with frequencies below 6 mHz (panel b). The same panel shows that the peaks in Pc5 activity and wave power response coincide with \dbxdt{} peaks observed on the ground by THEMIS/GMAG stations (Figure \ref{394}d) and peaks of geographic areas with \dbdt{} surpassing the 1.5 and 5 nT/s thresholds (Figure \ref{areas}a). However, the wave response around the substorm onset on 20 September 2015 is remarkably different. Panels c and d show very weak Pc5 and nearly nonexistent wave power response prior to the substorm onset. After 0703 UT, some Pc5 pulsation activity is seen in panel c, but wave power will only be enhanced 10 minutes later, at 0713 UT. This enhancement in wave power coincides with the maximum area with \dbdt{} $>$ 1.5 nT/s (Figure \ref{areas}, bottom panel) and the second particle injection observed by LANL (Figure \ref{lanl}, bottom row). Although the results of wave activity on 5 April 2010 seem to agree with the results reported by \citeA{Ngwira2018c}, the results observed during the 20 September 2015 are quite different. The later and weaker wave activity during the latter event are consistent with results showing that inclined shocks trigger less intense substorm activity \cite{Oliveira2014b,Oliveira2015a} and ULF wave activity \cite{Oliveira2020d} in comparison to shocks with small impact angles. \par

\section{Further shock inclination and strength evidence provided by MMS}

\begin{figure}[t]
		\centering
		\includegraphics[width=14cm]{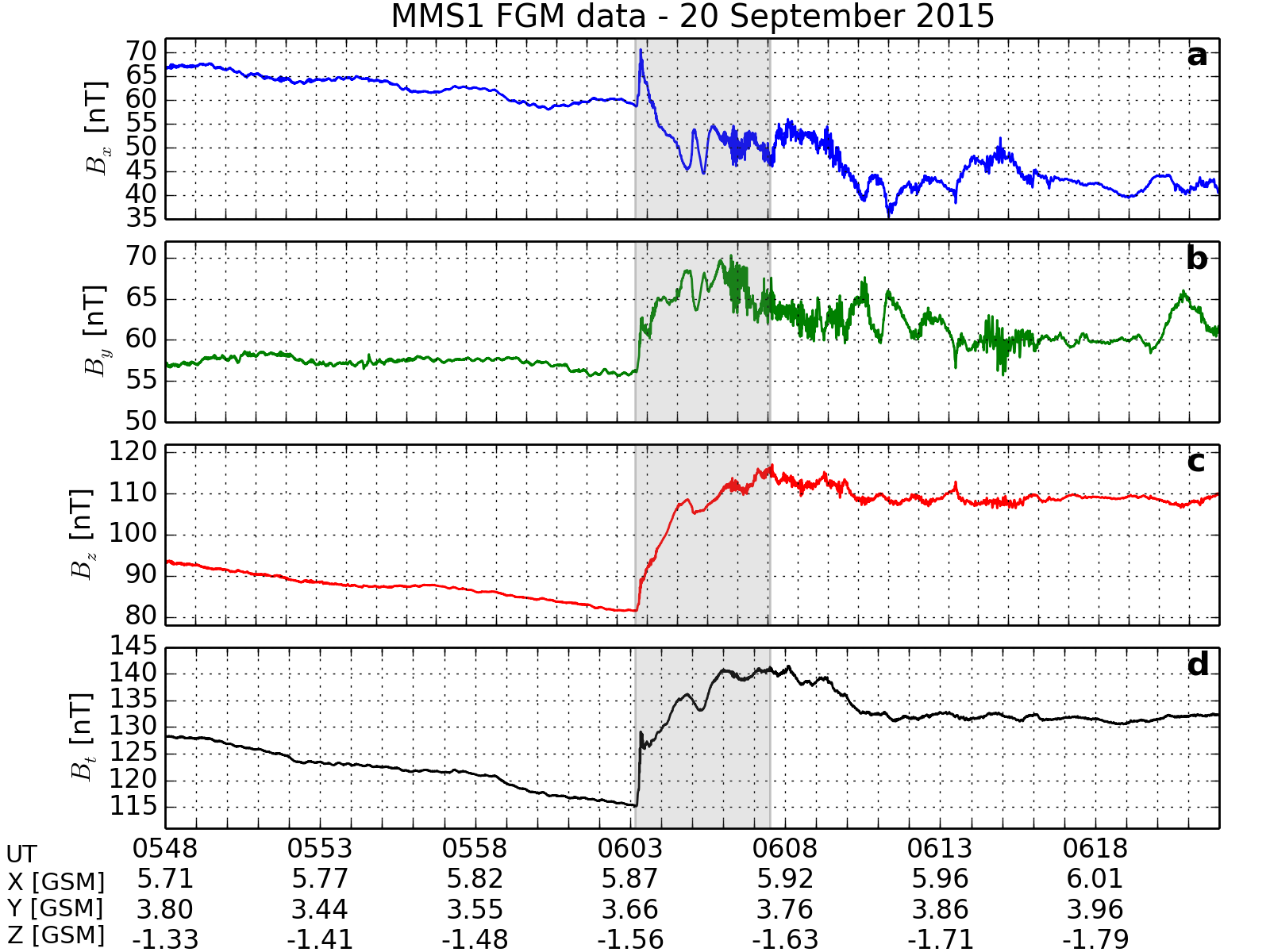}
		\caption{FGM magnetic field data collected by MMS1 on 20 September 2015. (a) $B_x$ component; (b) $B_y$ component; (c) $B_z$ component; and (d) magnetic field magnitude. A 1-second filter was applied to the data. The grey shaded areas span approximately 4 minutes and 44 seconds, which coincides very closely with the rise time indicated by the SMR index for the HIS (Figure \ref{522}a).}
		\label{mms_si}
	\end{figure}

		Since MMS was launched in March 2015, there are no data for the NFS on 5 April 2010. On the other hand, there are only magnetic field data available recorded around the onset of the HIS at 0603 UT. As shown in Figure \ref{orbits}, MMS was near $\sim$6\RE{} approximately in the path of the HIS. Figure \ref{mms_si} shows the three components of the magnetic field and its magnitude recorded by FGM/MMS1 in GSM coordinates. A 1-second smoothing filter was applied to the data. The occurrence of a sudden impulse is clearly seen a few seconds after 0603 UT. The shaded area in the figure shows a rise time of 4 minutes and 44 seconds, in great agreement with the rise time of 5 minutes observed after the HIS impact by SuperMAG stations (Figure \ref{522}a) and previous simulations and observations concerning very inclined shocks \cite{Oliveira2018a}. This is particularly evident in the $B_z$ component, which is usually the strongest component near geosynchronous orbit \cite<e.g., >[]{Wang2009}.

		Minimum variation analysis \cite{Sonnerup1998,Cameron2019a} applied to the magnetic field data collected by all four MMS spacecraft at shock impact yields \thxn = 141.09$^\circ$ $\pm$ 9.2$^\circ$, a very close value to the impact angle computed from Wind observations in GSM coordinates (143.2$^\circ$, Table \ref{table1}). Therefore, these observations by MMS from the magnetosphere show that the compression waves following the HIS impact and the HIS impact angle computed with Wind data from the solar wind are quite similar. These results agree with previous observations reported by \citeA{Oliveira2020d}. \par

\begin{figure}[t]
			\centering
			\includegraphics[width=14cm]{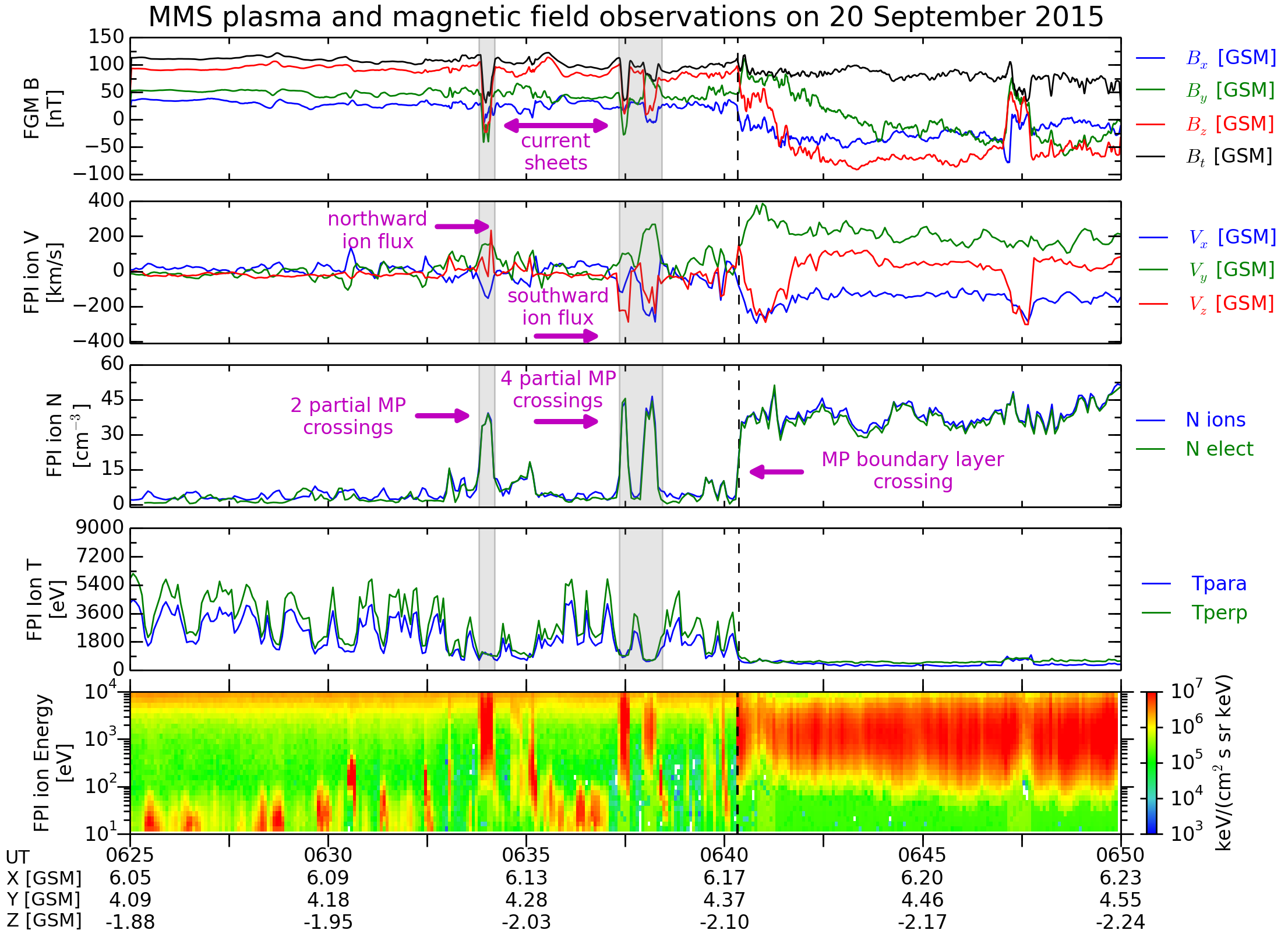}
			\caption{FGM magnetic field and FPI ion/electron observations by MMS during the highly inclined shock of 20 September 2015. The partial magnetopause crossings and the definitive magnetopause boundary layer crossing are show by the magenta arrows.}
			\label{mms}
		\end{figure}

		Approximately 20 minutes after the sudden impulse observation, MMS began to collect plasma data. Figure \ref{mms} shows 25 minutes of MMS FGM and FPI data following 0625 UT. From top to bottom, the figure shows all 3 components of the magnetic field data and its magnitude in GSM coordinates; three components of FPI ion velocity in GSM coordinates; ion and electron densities and temperatures; and the ion energy spectrogram.

		The figure shows that MMS observed two ion and electron density enhancements associated with a single pulse (near 0633 UT) and a double pulse (0637-0638 UT). This is probably due to two partial magnetopause crossings performed by MMS, and the structures are most likely two-dimensional current sheets \cite{Dong2018}. This is probably due to a dynamic pressure enhancement observed by Wind in the solar wind (Figure \ref{shocks}). All magnetic field components sharply decrease and increase after and during the pulses, and the same occurred with the x component of ion velocity, indicating enhanced ion fluxes toward the Earth. The first partial magnetopause crossing shows northward ion flux ($V_z$ $>$ 0), while the other two partial magnetopause crossings show southward ion flux ($V_z$ $<$ 0). These periods are also marked on the bottom panel by the regions with enhanced ion energy flux represented by the first narrow region (single pulse) and the following two narrow regions (double pulse). Later, at 0640:21 UT, MMS definitely crossed the magnetopause boundary entering the magnetosheath, as is clearly shown by the polarity inversion of $B_z$, increase in plasma density and decrease in plasma temperature, and enhanced ion energy fluxes, which are well-known magnetopause crossing signatures \cite<e.g.,>[]{Dong2018,Oliveira2020d}. Minimum variation analysis with the four MMS spacecraft yields \thxn{} = 137.67$^\circ$ $\pm$ 2.2$^\circ$ for the first structure, and \thxn{} = 143.74$^\circ$ $\pm$ 0.53$^\circ$ for the other two structures. These results also agree with the \thxn{} angle computed with magnetic field data collected by MMS in the magnetosheath: \thxn{} = 141.67$^\circ$ $\pm$ 6.8$^\circ$. \par

		Since \citeA{Collier2007} showed that subsequent pressure pulses and shocks following strong IP shocks usually have propagation directions nearly aligned with the direction propagation of the original shock, these minimal variation analyses indicate that the HIS was indeed very inclined. Additionally, MMS crossed the magnetopause boundary at $X_{GSM}$ $\sim$6.17 \RE, within geosynchronous orbit, indicting a very strong shock impact. This indicates that the magnetosphere was under a very high pressure state prior to substorm triggering at 0703 UT. Therefore, these MMS observations support Wind's observations of a very strong and inclined shock in the solar wind (Figures \ref{shocks} and \ref{standoff}).

\section{Discussion and conclusion}

	In this study, we used the methodology suggested by \citeA{Ngwira2018c} to investigate the magnetosphere-ionosphere response and ground magnetic field response during shock-induced substorm activity by taking a space-borne and ground-based, multi-instrument approach. By carefully analyzing two isolated substorms triggered by similarly strong shocks with very different inclinations, we found that the general geomagnetic activity represented by geomagnetic indices, energetic particle injection, auroral poleward expansion, westward auroral electrojet current, ground \dbdt{} variations, and ULF wave activity following the NFS were overall higher than the same effects following the HIS. These findings are supported by many experimental \cite{Takeuchi2002b,Wang2006a,Oliveira2015a,Oliveira2016a,Oliveira2018b,Shi2019b,Oliveira2020d} and simulation \cite{Guo2005,Samsonov2011a,Oliveira2014b,Samsonov2015,Selvakumaran2017} works in the literature. \par

	However, when conducting simulations, it is a simple task to control for IMF and plasma parameters before shock impacts to account for preconditioning effects \cite{Zhou2001}. For example, \citeA{Oliveira2014b} kept IMF $B_z$ slightly negative before the impacts of an inclined shock and a purely frontal shock in their simulations. The authors launched the shocks in the simulations with step-like jumps in IMF and plasma parameters, meaning that the level of compression and IMF $B_z$ intensity are closely controlled in the simulations. However, that is not always the case in the real solar wind because solar wind discontinuities and other solar wind structures may follow IP shocks \cite{Tsurutani2011a} with very close propagation directions \cite{Collier2007}. As seen in Figure \ref{shocks}, the y and z components of the IMF and plasma velocity are much more variable in the HIS case, indicating that the propagation direction of the solar wind structures closely followed the original shock impact angle of \thxn{} $\sim$143$^\circ$. Additionally, MMS observed three partial magnetopause crossings with current sheet normal angles nearly aligned with \thxn{} (Figure \ref{mms}), supporting the many effects observed in space and on the ground related to the large shock impact angle of the HIS. \par

	As shown in Table \ref{table2}, the density ratio ($X_\rho$) and the dynamic pressure ratio ($X_{P_d}$) are higher for the NFS compared to the HIS (2.70 and 1.71, and 4.31 and 2.55 respectively), while the magnetosonic Mach number is slightly higher for the HIS. Such differences typically arise when frontal and inclined shocks are compared. \citeA{Samsonov2011a} calculated the solar wind velocity ($v_x$ and $v_y$) and the dynamic pressure downstream of IP shocks with different inclination angles using the Rankine-Hugoniot conditions and showed that $v_x$ and dynamic pressure grow higher for nearly frontal shocks (see his Figure 2 and Table 1). This effect results in stronger magnetospheric compression not only in the dayside region but also in the magnetotail. Additionally, higher $v_x$ (and consequently higher electric field) downstream nearly frontal shocks results in stronger solar wind driving of magnetospheric activity including the \dbdt{} variations addressed in this study. Therefore, the differences between the \dbdt{} variations caused by the NFS and HIS discussed in this study should be expected and are supported by many experimental and modeling studies \cite{Oliveira2018a}. \par

	Geospace observations by THEMIS-A and LANL-1989-046 of energetic particle injections on 5 April 2010 in the post-midnight sector were shortly followed by intense \dbdt{} variation observed by MCGR, GAKO, FYKN, WHIT, and INUV (5 minutes later). All variations had $|$\dbxdt$|$ $>$ 5 nT/s, hence surpassing a critical \dbdt{} value linked to the generation of dangerous GICs \cite{Molinski2000,Weygand2021}. During the same event, intense auroral brightening was observed by all stations (except for INUV later), confirming previous observations by \citeA{Ngwira2018c}. This auroral expansion is consistent with the intensification of the westward electrojet current known as a direct manifestation of the substorm current wedge \cite{McPherron1972}. The substorm current wedge is a large-scale current structure that enters the ionosphere in the eastern edge due to the explosive release of substorm-time magnetotail energy and returns to space from the ionospheric western edge as a result of current closure \cite{McPherron1972,Pytte1976}. The results of this work then show that a substorm triggered during times of very high and symmetric compression conditions on 5 April 2010 may have directly contributed to the efficiency and rapidness of this energy release and intensification of the substorm current wedge, ultimately leading to intense auroral brightening and \dbdt{} variations observed on the ground. Conversely, these enhanced symmetric conditions did not occur during the 20 September 2015 event; quite the contrary, the asymmetric and weaker magnetospheric compression state during the substorm triggering led to very weak energetic particle injection levels observed by THEMIS-E, and a double relatively weaker and slower injection being observed by LANL-01A. These asymmetric effects, related to the original and high shock impact angle, did not cause a clear poleward expansion of the auroral oval hence leading to very mild ground \dbxdt{} variations observed by FSMI, ATHA, and INUV with $|$\dbxdt$|$ $<$ 1.5 nT/s after substorm onset, with WHIT later observing $|$\dbxdt$|$ $>$ 5 nT/s probably associated with the double injection peak observed by LANL in space. Table \ref{table3} shows a chronological sequence of similar events occurring during the two dates for a general comparison. \par

	\begin{figure}[t]
		\centering
		\includegraphics[width=13cm]{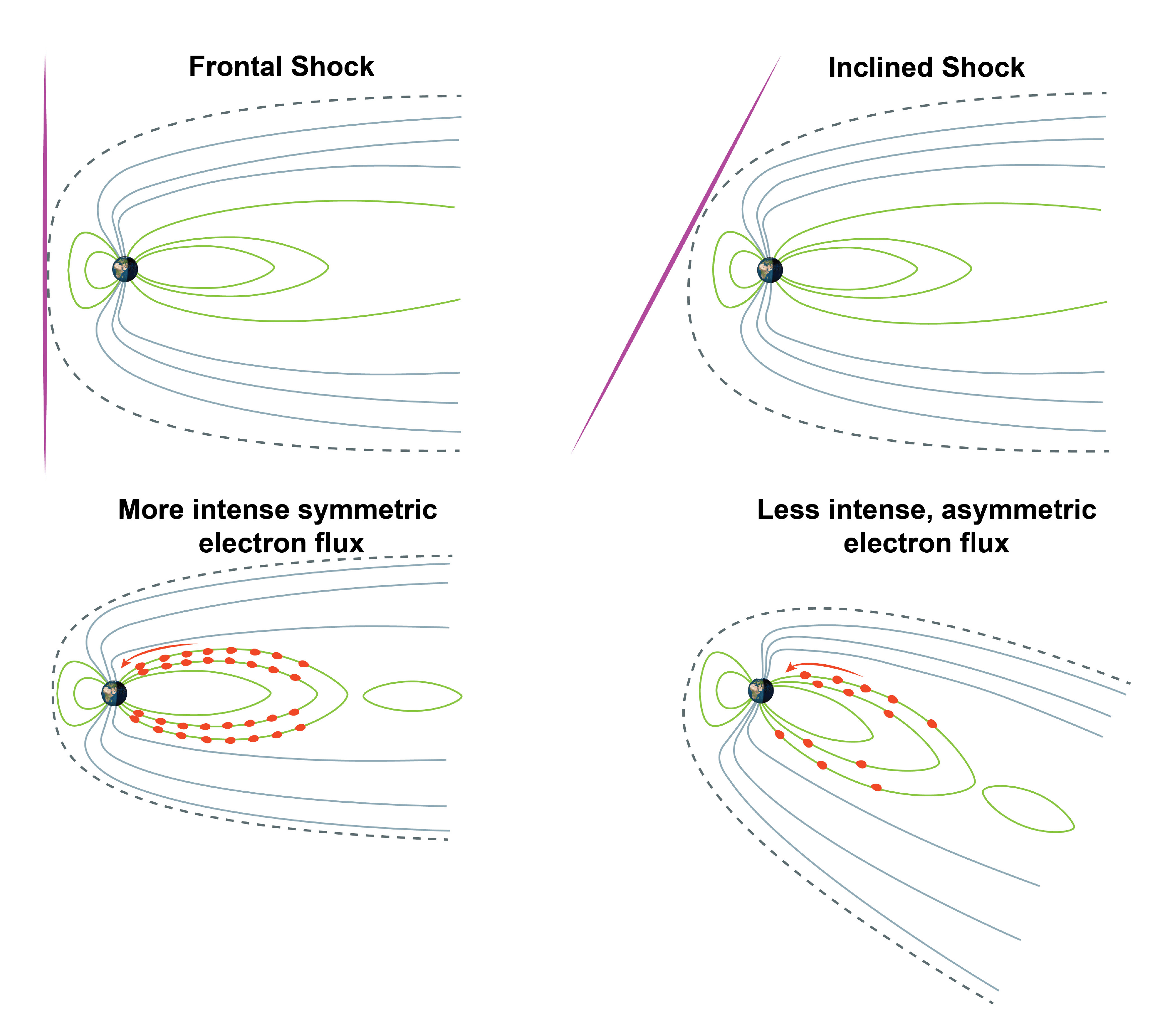}
		\caption{Schematic representation of the magnetosphere and its response to shocks with different orientations.}
		\label{sketch}
	\end{figure}

	The minimum magnetopause positions observed for both shocks and shown in Figure \ref{standoff} are comparable to results of numerical simulations for an extremely fast CME simulated by \citeA{Welling2021} ($\sim$4.8\RE), and for a Carrington-like CME simulated by \citeA{Blake2021b} ($\sim6.3$\RE). Our results also compare with the minimum magnetopause standoff positions caused by most likely fast CMEs and obtained from historical magnetograms in September 1909 \cite{Love2019b} ($\sim$5.9\RE), and February 1941 \cite{Hayakawa2021a} ($\sim$7.2\RE), which were all followed by intense ground magnetic field intensifications. Additionally, Figure \ref{mms} shows that MMS definitely moved from the magnetosphere into the magnetosheath at $X$ $\sim$6.18\RE{} approximately 24 minutes before the substorm onset on 20 September 2015 at 0703 UT, shortly following two partial magnetopause crossings where the current sheets had normal vectors almost aligned with the original shock propagation direction. These MMS observations indicate that the magnetosphere was very strongly and asymmetrically compressed during the September 2015 substorm. Therefore, these simulations and observational results support the observations indicating that the HIS was indeed a very strong shock. \par

	The very intense substorm of 5 April 2010 has been extensively studied before \cite<e.g.,>[]{McComas2012,Lotoaniu2015,Nishimura2020}. For example, \citeA{Nishimura2020} reported on the occurrence of very intense magnetosphere-ionosphere-thermosphere response during high levels of magnetospheric pressure and highly depressed levels of the southward component of the IMF represented by total electron content, conductance, neutral winds, and auroral brighting and electrojets. The authors noted that such enhancements were much more intense than those occurring during classic substorms. On the space weather front, the Galaxy 15 satellite experienced an anomaly on 5 April 2010 at 0948 UT when orbiting Earth around the magnetic midnight sector. Intelsat S.A, the company that owed Galaxy 15, completely lost control of the satellite, which remained a ``zombie" satellite until 18 October 2010 when contact with the satellite was eventually reestablished \cite{Lotoaniu2015}. \citeA{Lotoaniu2015} suggested that Galaxy 15 was in a very hostile environment with extreme energetic particle and magnetic field conditions on 5 April 2010, being in ``the wrong place at a critical time" \cite{Allen2010}. However, to our knowledge, assuming there were satellites in geosynchronous orbit around midnight, there have been no publications reporting on extreme space weather impacts on geosynchronous satellites during the 20 September 2015 event, even though the substorm on that day occurred during a very intense compression state of the magnetosphere. We suggest that the geoeffectiveness of substorms triggered during very high and symmetric magnetospheric compressions will be very intense since energy will be rapidly and explosively released in a very narrow region in space, and more work is needed to understand these effects including the 2015 event. \par

	Our results clearly show that \dbdt{} variations can easily surpass the 1.5 nT/s threshold during the isolated substorms triggered by the shocks here investigated. However, thresholds of $|$\dbdt$|$ $>$ 5 nT/s were seldom observed during the HIS-triggered substorm (4 times), whereas this threshold surpassed 5 nT/s during the NFS-triggered substorm multiple times. Although \citeA{Weygand2021} already showed \dbdt{} surpassing these values during substorms, this present work clearly indicates that the shock impact angle can be a crucial factor controlling the ground \dbdt{} response. Global \dbdt{} maps in North America show that the overall geographic areas with \dbdt{} surpassing the thresholds 1.5 and 5 nT/s were generally higher in the 5 April 2010 event, with intense ULF/Pc5 wave activity occurring at times of maximum areas in both events, but with wave actitiy being higher in the NFS case compared to the HIS case. Considering that these \dbdt{} thresholds can cause damage to power grid equipment operating in different time scales \cite{Molinski2000,Pulkkinen2013,Oliveira2018b}, power plant operators can take advantage of these results to plan ahead and avoid long-term damage associated with space weather effects particularly caused by IP shocks that are observed at L1 and bound to impact Earth nearly head-on \cite{Vorotnikov2011,Paulson2012,Kruparova2013}. \par 

	\citeA{Tsurutani2021} used over two years of GIC data collected in southern Finland to show that 76\% of the events with GIC $>$ 30 A were caused by supersubstorms, whereas all events with GIC $>$ 10 A were caused by intense substorms. \citeA{Tsurutani2021} noted that all events showed shock-related GIC effects. \citeA{Hajra2018} noted that supersubstorms are usually triggered by IP shocks strongly compressing the magnetosphere under extremely high solar wind ram pressure, associated with intense southward turnings of the IMF $B_z$ component. Therefore, we suggest that the nearly symmetric compression of the magnetosphere is a crucial factor for the triggering of supersubstorms due to the subsequent intensity amplification of the resulting substorm current wedge \cite{McPherron1972,Pytte1976}. Therefore, statistical studies connecting shock-induced supersubstorms with the shock impact angle are needed to address this open question. \par

	The results of this work also have further implications to inter-hemispheric studies of magnetosphere-ionosphere-thermosphere response. \citeA{Xu2020a} showed that shock impact angle inclinations with respect to the solar magnetic equatorial plane strongly controls the timing and intensity of conjugated ground \dbdt{} responses, with the hemisphere first receiving the impact responding first with more intense \dbdt{} variations. As shown in Figure \ref{sketch}, symmetric magnetospheric compressions can enhance more intense and symmetric electron fluxes (left column), whereas shocks with north-south impact angle inclinations can lead to less intense and more asymmetric electron fluxes, with the more intense electron fluxes occurring in the hemisphere that first receives the shock impact (right column). Therefore, further studies involving energetic particle fluxes, ionospheric conductance, field-aligned currents, total electron content, neutral wind and mass density occurring during substorms triggered during asymmetric conditions are necessary to improve our understanding of the shock impact angle control of such geomagnetic activity. \par

	\begin{table}
		\centering
		\begin{tabular}{l l c l l}
			\hline
				Date & \hspace*{1cm}5 April 2010 & & Date & \hspace*{1cm}20 September 2015 \\
				UT & \hspace*{1cm}Event description & & UT & \hspace*{1cm}Event description \\
			\hline
				& & & & \\
				0826 & \begin{minipage}[t]{0.42\textwidth}Nearly frontal shock impact\end{minipage} & &
				0603 & \begin{minipage}[t]{0.42\textwidth}Highly inclined shock impact\end{minipage} \\
				& & & & \\
				0828 & \begin{minipage}[t]{0.42\textwidth}Strong SI$^+$ event occurrence (SMR amplitude $\sim$30 nT) with rise time ($\sim$2 minute)\end{minipage} & & 
				0603 & \begin{minipage}[t]{0.42\textwidth}Strong SI$^+$ event occurrence (SMR amplitude $\sim$30 nT) with rise time ($\sim$5 minute)\end{minipage} \\ 
				& & & & \\
				0828 & \begin{minipage}[t]{0.42\textwidth}Minimum magnetopause standoff position (\XMP{} = 5.4\RE)\end{minipage} & & 
				0608 & \begin{minipage}[t]{0.42\textwidth}Minimum magnetopause standoff position (\XMP{} = 6.4\RE)\end{minipage} \\
				& & & & \\
				& \begin{minipage}[t]{0.42\textwidth}{}\end{minipage} & & 
				0637 & \begin{minipage}[t]{0.42\textwidth}MMS observes a partial magnetopause crossing and another two 4 minutes later, with current sheet normal vectors almost aligned with the HIS propagation direction \end{minipage} \\
				& & & & \\
				& \begin{minipage}[t]{0.42\textwidth}{}\end{minipage} & & 
				0641 & \begin{minipage}[t]{0.42\textwidth}MMS definitely enters the magnetosheath at 6.18\RE\end{minipage} \\
				& & & & \\
				0903 & \begin{minipage}[t]{0.42\textwidth}Substorm onset 37 minutes after shock impact\end{minipage} & & 
				0703 & \begin{minipage}[t]{0.42\textwidth}Substorm onset 60 minutes after shock impact\end{minipage} \\
				& & & & \\
				0903 & \begin{minipage}[t]{0.42\textwidth}Extreme \dbxdt{} peak (--22.6 nT/s) recorded by Gakona\end{minipage} & & 
				0705 & \begin{minipage}[t]{0.42\textwidth}Extreme \dbxdt{} peak (-- 7 nT/s) recorded by White Horse\end{minipage} \\
				& & & & \\
				0903 & \begin{minipage}[t]{0.42\textwidth}Very intense regional SMU index enhancements in the afternoon sector\end{minipage} & & 
				0705 & \begin{minipage}[t]{0.42\textwidth}Intense regional SMU index enhancements in the afternoon sector\end{minipage} \\
				& & & & \\
				0903 & \begin{minipage}[t]{0.42\textwidth} Intense electron flux enhancements observed at geosynchronous orbit by LANL-1989-046\end{minipage} & & 
				0705 & \begin{minipage}[t]{0.42\textwidth} Intense electron energy injection observed at geosynchronous orbit by LANL-01A\end{minipage} \\
				& & & & \\
				     & \begin{minipage}[t]{0.42\textwidth} \end{minipage} & & 
				0713 & \begin{minipage}[t]{0.42\textwidth} Second intense electron flux enhancements observed at geosynchronous orbit by LANL-01A\end{minipage} \\
				& & & & \\
				0903 & \begin{minipage}[t]{0.42\textwidth}Maximum geographic area coverage with SECS \dbdt{} surpassing 1.5 nT/s \end{minipage} & & 
				0713 & \begin{minipage}[t]{0.42\textwidth}Maximum geographic area coverage with SECS \dbdt{} surpassing 1.5 nT/s \end{minipage} \\
				& & & & \\
				0909 & \begin{minipage}[t]{0.40\textwidth} Intense electron fluxes and electric fields observed by THEMIS-A\end{minipage} & & 
				0703 & \begin{minipage}[t]{0.40\textwidth} Weak electron fluxes and electric fields observed by THEMIS-E\end{minipage} \\
				& & & & \\
				0903 & \begin{minipage}[t]{0.42\textwidth}Maximum ULF/Pc5 wave activity observed by White Horse \end{minipage} & & 
				0713 & \begin{minipage}[t]{0.42\textwidth}Onset of intense ULF/Pc5 wave activity observed by Fort Simpson\end{minipage} \\
				& & & & \\
				0929 & \begin{minipage}[t]{0.42\textwidth}{Minimum SuperMAG westward auroral electrojet index (SML = --2351 nT)}\end{minipage} & & 
				0708 & \begin{minipage}[t]{0.42\textwidth}Minimum SuperMAG westward auroral electrojet index (SML = --1831 nT)\end{minipage} \\
			\hline
		\end{tabular}
		\caption{Summary of the chain of events occurring on 5 April 2010 (left part) and 20 September 2015 (right part).}
		\label{table3}
	\end{table}

\section*{Acknowledgments}

	DMO and EZ thank the NASA Space Weather Science Applications Operations 2 Research grant No. 20-SWO2R20-2-0014. JMW acknowledges the following funding: NASA THEMIS contract SA3650-26326:44, NASA grant number 80NSSC18K1220, NASA HPDE contract 80GSFC17C0018, and NASA grant number 80NSSC18K1227. MDH was supported by NASA 80NSSC19K0907.

\section*{Data Availability Statement}

	The ACE and Wind solar wind and IMF data are available from the NASA/OMNI website \url{https://omniweb.gsfc.nasa.gov}. The SuperMAG ground magnetometer data are available at \url{https://supermag.jhuapl.edu} at the tab ``Indices". THEMIS space-borne (magnetic field and plasma) and ground-based magnetometer and image data (GMAG and ASI) are available at \url{http://themis.ssl.berkeley.edu/data/themis/}. The LANL data are available at the Zenodo repository (\url{https://zenodo.org/record/5532746#.YVIM5S2cbFw}). The SECS data are available to the public through the Virtual Magnetospheric Observatory (VMO) (\url{http://vmo.igpp.ucla.edu/data1/SECS}). MMS FGM and FPI data were downloaded from the MMS Science Data Center website (\url{https://lasp.colorado.edu/mms/sdc/public/}) provided by the Laboratory for Atmospheric and Space Physics (LASP), University of Colorado, Boulder.

\newpage

\end{document}